\documentclass[article,nojss]{jss}

\usepackage{thumbpdf,lmodern}

\usepackage{framed}
\usepackage{listings} 
\usepackage{natbib}

\usepackage{booktabs} 
\usepackage{graphicx} 
\usepackage{caption}
\usepackage{subcaption} 
\usepackage{bm}
\usepackage{amsfonts}
\usepackage{stackengine}
\usepackage{amsmath}
\usepackage{amssymb}
\usepackage{mathrsfs}
\usepackage{bbm}
\usepackage{tikz}
\usetikzlibrary{calc}
\usepackage{tabularx}
    \newcolumntype{L}{>{\raggedright\arraybackslash}X}

\author{Zhi Zhao\\University of Oslo 
   \And Marco Banterle\\London School of Hygiene\\ \& Tropical Medicine
   \And Leonardo Bottolo\\University of Cambridge\\Alan Turing Institute
   \AND Sylvia Richardson\\University of Cambridge\\Alan Turing Institute
   \And Alex Lewin$^*$\\London School of Hygiene\\ \& Tropical Medicine
   \And Manuela Zucknick$^*$\\University of Oslo}
\Plainauthor{Zhao Z, Banterle M, Bottolo L, Richardson S, Lewin A and Zucknick M}

\title{\pkg{BayesSUR}: An \proglang{R} package for high-dimensional multivariate Bayesian variable and covariance selection in linear regression}
\Plaintitle{BayesSUR: An R package package for high-dimensional multivariate Bayesian variable and covariance selection in linear regression}
\Shorttitle{\pkg{BayesSUR}}

\Abstract{
 In molecular biology, advances in high-throughput technologies have made it possible to study complex multivariate phenotypes and their simultaneous associations with high-dimensional genomic and other omics data, a problem that can be studied with high-dimensional multi-response regression, where the response variables are potentially highly correlated.

To this purpose, we recently introduced several multivariate Bayesian variable and covariance selection models, e.g., Bayesian estimation methods for sparse seemingly unrelated regression for variable and covariance selection. Several variable selection priors have been implemented in this context, in particular the hotspot detection prior for latent variable inclusion indicators, which results in sparse variable selection for associations between predictors and multiple phenotypes. We also propose an alternative, which uses a Markov random field (MRF) prior for incorporating prior knowledge about the dependence structure of the inclusion indicators. Inference of Bayesian seemingly unrelated regression (SUR) by Markov chain Monte Carlo methods is made computationally feasible by factorisation of the covariance matrix amongst the response variables.

In this paper we present \pkg{BayesSUR}, an \proglang{R} package, which allows the user to easily specify and run a range of different Bayesian SUR models, which have been implemented in \proglang{C++} for computational efficiency. The \proglang{R} package allows the specification of the models in a modular way, where the user chooses the priors for variable selection and for covariance selection separately. We demonstrate the performance of sparse SUR models with the hotspot prior and spike-and-slab MRF prior on synthetic and real data sets representing eQTL or mQTL studies and \emph{in vitro} anti-cancer drug screening studies as examples for typical applications.

}

\Keywords{Seemingly unrelated regression, Bayesian multivariate regression, structured covariance matrix, Markov random field prior, multi-omics data}
\Plainkeywords{Seemingly unrelated regression, Bayesian multivariate regression, structured covariance matrix, Markov random field prior, multi-omics data}

\Address{
Zhi Zhao, Manuela Zucknick\\
Oslo Centre for Biostatistics and Epidemiology\\
Department of Biostatistics\\
Institute of Basic Medical Sciences\\
University of Oslo\\
P.O. Box 1122 Blindern\\
0317 Oslo, Norway\\
E-mail: \email{zhi.zhao@medisin.uio.no}, \email{manuela.zucknick@medisin.uio.no}\\

Marco Banterle, Alex Lewin\\
Department of Medical Statistics\\
Faculty of Epidemiology and Population Health\\
London School of Hygiene \& Tropical Medicine\\
Keppel St, Bloomsbury\\
London WC1E 7HT, United Kingdom\\
E-mail: \email{marco.banterle@gmail.com}, \email{alex.lewin@lshtm.ac.uk}\\

Leonardo Bottolo\\
Department of Medical Genetics\\
University of Cambridge\\
J. J. Thomson Avenue\\
Cambridge CB2 0QQ, United Kingdom\\
E-mail: \email{lb664@cam.ac.uk}\\

Sylvia Richardson\\
MRC Biostatistics Unit\\
University of Cambridge\\
Robinson Way\\
Cambridge CB2 0SR, United Kingdom\\
E-mail: \email{sylvia.richardson@mrc-bsu.cam.ac.uk}
}

\begin{document}



\clearpage
\section{Introduction} \label{sec:intro}

With the development of high-throughput technologies in molecular biology, the large-scale molecular characterisation of biological samples has become common-place, for example by genome-wide measurement of gene expression, single nucleotide polymorphisms (SNP) or CpG methylation status. Other complex phenotypes, for example, pharmacological profiling from large-scale cancer drug screens, are also measured in order to guide personalised cancer therapies \citep{Garnett2012, Barretina2012, Gray2015}. 
The analysis of joint associations between multiple correlated phenotypes and high-dimensional molecular features is challenging.  

When multiple phenotypes and high-dimensional genomic information are jointly analysed, the Bayesian framework allows to specify in a flexible manner the complex relationships between the highly structured data sets. Much work has been done in this area in recent years. Our software package \pkg{BayesSUR} gathers together several models that we have proposed for high-dimensional regression of multiple responses and also introduces a novel model, allowing for different priors for variable selection in the regression models and for different assumptions about the dependence structure between responses.

Bayesian variable selection uses latent indicator variables to explicitly add or remove predictors in each regression during the model search. Here, as we consider simultaneously many predictors and several responses, we have a matrix of variable selection indicators. Different variable selection priors have been proposed in the literature. For example, \cite{Jia2007} mapped multiple phenotypes to genetic markers (i.e., expression quantitative trait loci, eQTL) using the spike-and-slab prior and hyper predictor-effect prior. \cite{Liquet2017} incorporated group structures of multiple predictors via a (multivariate) spike-and-slab prior. The corresponding \proglang{R} package \pkg{MBSGS} is available on CRAN (\url{https://cran.r-project.org/packages=MBSGS}). \cite{Bottolo2011} and \cite{Lewin2016} further proposed the hotspot prior for variable selection in multivariate regression, in which the probability of association between the predictors and responses is decomposed multiplicatively into predictor and response random effects. This prior is implemented in a multivariate Bayesian hierarchical regression setup in the software \pkg{R2HESS}, available from \url{https://www.mrc-bsu.cam.ac.uk/software/}. \cite{Lee2017} used the Markov random field (MRF) prior to encourage joint selection of the same variable across several correlated response variables. Their \proglang{C}-based \proglang{R} package \pkg{mBvs} is available on CRAN (\url{https://CRAN.R-project.org/package=mBvs}).


For high-dimensional predictors and multivariate responses, the space of models is very large. To overcome the infeasibility of the enumerated model space for the MCMC samplers in the high-dimensional situation, \cite{Bottolo2010} proposed an Evolutionary Stochastic Search (ESS) algorithm based on Evolutionary Monte Carlo. This sampler has been extended in a number of situations and efficient implementation of ESS for multivariate Bayesian hierarchical regression has been provided by the \proglang{C++}-based \proglang{R} package \pkg{R2GUESS} (\url{ https://CRAN.R-project.org/package=R2GUESS}) \citep{Liquet2016}. 
\cite{Richardson2011} proposed a new model and computationally efficient hierarchical evolutionary stochastic search algorithm (HESS) for multi-response (i.e., multivariate) regression which assumes independence between residuals across responses and is implemented in the \pkg{R2HESS} package. \cite{Petretto2010} used the inverse Wishart prior on the covariance matrix of residuals in order to do simultaneous analysis of multiple response variables allowing for correlations in response residuals, for more moderate sized data sets. 

In order to analyse larger numbers of response variables, yet retain the ability to estimate dependence structures between them, sparsity can be introduced into the residual covariances, as well as into the regression model selection. \cite{Holmes2002} adapted seemingly unrelated regression (SUR) to the Bayesian framework and used a Markov chain Monte Carlo (MCMC) algorithm for the analytically intractable posterior inference. The hyper-inverse Wishart prior has been used to learn a sparser graph structure for the covariance matrix of high-dimensional variables \citep{Carvalho2007, Wang2010, Bhadra2013}, thus performing covariance selection. However, these approaches are not computationally feasible if the number of input variables is very large. \cite{Banterle2018} recently developed a Bayesian variable selection model which employs the hotspot prior for variable selection, learns a structured covariance matrix and implements the ESS algorithm in the SUR framework to further improve computational efficiency.

The \pkg{BayesSUR} package implements many of these possible choices for high-dimensional multi-response regressions by allowing the user to choose among three different prior structures for the residual covariance matrix and among three priors for the joint distribution of the variable selection indicators. This includes a novel model setup, where the MRF prior for incorporating prior knowledge about the dependence structure of the inclusion indicators is combined with Bayesian SUR models \citep{Zhao2021}. \pkg{BayesSUR} employs ESS as a basic variable selection algorithm.


\section{Models specification} \label{sec:models}

The \pkg{BayesSUR} package fits a Bayesian seemingly unrelated regression model with a number of options for variable selection, and where the covariance matrix structure is allowed to be diagonal, dense or sparse. It encompasses three classes of Bayesian multi-response linear regression models: Hierarchical Related Regressions (HRR, \cite{Richardson2011}), dense and Sparse Seemingly Unrelated Regressions (dSUR and SSUR, \cite{Banterle2018}), and the Structured Seemingly Unrelated Regression, which makes use of a Markov random field (MRF) prior \citep{Zhao2021}.

The regression model is written as
\begin{align*}\tag{1}\label{orginEq}
\mathbf{Y} &= \mathbf{X}\bm{B} + \mathbf{U}, \\
\text{vec}(\mathbf{U}) &\sim \mathcal{N}(\mathbf{0},\ C \otimes \mathbb{I}_n )
\end{align*}
where $\mathbf{Y}$ is a $n\times s$ matrix of outcome variables with $s \times s$ covariance matrix $C$, $\mathbf{X}$ is a $n\times p$ matrix of predictors for all outcomes and $\bm{B}$ is a $p\times s$ matrix of regression coefficients.

We use a binary latent indicator matrix $\bm\Gamma=\{\gamma_{jk}\}$ to perform variable selection. A spike-and-slab prior is used to find a sparse relevant subset of predictors that explain the variability of $\mathbf{Y}$: 
conditional on $\gamma_{jk} = 0$ ($j=1,\cdots,p$ and $k=1,\cdots,s$) we set $\beta_{jk} = 0$ and conditional on $\gamma_{jk} = 1$ regression coefficients follow a diffuse Normal:
\begin{align*}
\bm{\beta}_{\bm{\gamma}} \sim \mathcal{N}\left(\bm{0}, W_{\bm{\gamma}}^{-1} \right). \tag{2}\label{diffuseNormal}
\end{align*}
where $\bm{\beta}=\text{vec}(\bm{B})$, $\bm{\gamma}=\text{vec}(\Gamma)$, $\bm{\beta}_{\bm{\gamma}}$ consists of the selected regression coefficients only (i.e., where $\gamma_{jk}=1$), and likewise $W_{\bm{\gamma}}$ is the sub-matrix of $W$ formed by the 
corresponding selected coefficients.

The precision matrix $W$ is generally decomposed into a shrinkage coefficient 
and a matrix that governs the covariance structure of the regression coefficients. Here we use 
$W = w^{-1} \mathbb{I}_{sp}$, meaning that all the regression coefficients are \textit{a priori} independent, with an inverse gamma hyperprior on the shrinkage coefficient $w$, i.e., $w \sim \mathcal{IG}amma (a_w, b_w)$. The binary latent indicator matrix $\bm\Gamma$  has three possible options for priors: the independent hierarchical Bernoulli prior, the hotspot prior and the MRF prior. The covariance matrix $C$ also has three possible options for priors: the independent inverse gamma prior, the inverse Wishart prior and hyper-inverse Wishart prior. Thus, we consider nine possible models (Table \ref{tab:modelList}) across all combinations of three priors for $C$ and three priors for $\bm{\Gamma}$.

\begin{table}[tph]
\centering
\begin{tabular}{l c c c } 
\hline
                 & $\gamma_{jk} \sim$ Bernoulli  & $\gamma_{jk} \sim$ Hotspot  & $\bm{\gamma} \sim$ MRF  \\ 
\hline
$C \sim$ indep   & HRR-B                        &  HRR-H                     &  HRR-M          \\
$C \sim \mathcal{IW}$    & dSUR-B                        &  dSUR-H                     &  dSUR-M          \\
$C \sim \mathcal{HIW}_{\mathcal{G}}$   & SSUR-B                        &  SSUR-H                     &  SSUR-M          \\
\hline
\end{tabular}
\caption{ Nine models across three priors of $C$ by three priors of $\bm\Gamma$  }
\label{tab:modelList}
\end{table}

\subsection{Hierarchical Related Regression (HRR)} \label{sec:modelHRR}

The Hierarchical Related Regression model assumes that $C$ is a diagonal matrix
\[
C = \begin{pmatrix}
\sigma_1^2 & \cdots & 0 \\
          & \ddots &  \\
0          & \cdots & \sigma_s^2 
\end{pmatrix}, \tag{3}\label{diagC}
\]
which translates into conditional independence between the multiple response variables, so the likelihood factorises across responses. 
An inverse gamma prior is specified for the residual covariance, i.e., $\sigma_k^2 \sim \mathcal{IG}amma(a_{\sigma},b_{\sigma})$ which, combined with the priors in (\ref{diffuseNormal}) is conjugate with the likelihood in the model in (\ref{orginEq}).
We can thus sample the variable selection structure $\bm\Gamma$  marginally with respect to $C$ and $\bm{B}$. For inference for this model, \cite{Richardson2011} implemented the hierarchical evolutionary stochastic search algorithm (HESS).

\smallskip
\subsubsection{2.1.1. HRR with independent Bernoulli prior} \label{sec:modelHRRind}

For a simple independent prior on the regression model selection, 
the binary latent indicators follow a Bernoulli prior
\begin{equation*}
\gamma_{jk} |\omega_{jk} \sim \mathcal{B}er (\omega_j), \ \ j=1,\cdots,p, \ k=1,\cdots,s,  \tag{4}\label{BernoulliPrior}
\end{equation*}
with a further hierarchical Beta prior on $\omega_j$, i.e., $\omega_j \sim \mathcal{B}eta (a_{\omega},b_{\omega})$, which quantifies the probability for each predictor to be associated with any response variable.

\smallskip
\subsubsection{2.1.2. HRR with hotspot prior}

\cite{Richardson2011} and \cite{Bottolo2011} proposed decomposing the probability of association parameter $\omega_{jk}$ in (\ref{BernoulliPrior}) as $\omega_{jk} = o_k \times \pi_j$, where $o_k$ accounts for the sparsity of each response model and $\pi_j$ controls the propensity of each predictor to be associated with multiple responses simultaneously.
\begin{align*} \tag{5}\label{hotspotPrior}
\gamma_{jk} |\omega_{jk} &\sim \mathcal{B}er (\omega_{jk}), \ \ j=1,\cdots,p, \ k=1,\cdots,s,  \\
\omega_{jk} &= o_k \times \pi_j , \\
o_k &\sim \mathcal{B}eta (a_{o},b_{o}), \\
\pi_j &\sim \mathcal{G}amma(a_{\pi},b_{\pi}).
\end{align*}

\smallskip
\subsubsection{2.1.3. HRR with MRF prior}

To consider the relationship between different predictors and associate highly correlated responses with the same predictors, we set a Markov random field prior on the latent binary vector $\bm{\gamma}$
\begin{align*}
f(\bm{\gamma} |d,e,G) &\propto \exp\{d\mathbf 1^\top\bm{\gamma} + e \bm{\gamma}^\top G\bm{\gamma}\}, \tag{6}\label{MRFprior}
\end{align*}
where $G$ is an adjacency matrix containing prior information about similarities amongst the binary model selection indicators $\bm{\gamma}=\text{vec}(\bm\Gamma)$.
The parameters $d$ and $e$ are treated as fixed in the model. Alternative approaches include the use of a hyperprior on $e$ \citep{Stingo2011} or to fit the model repeatedly over a grid of values for these parameters, in order to detect the phase transition boundary for $e$ \citep{Lee2017} and to identify a sensible combination of $d$ and $e$ that corresponds to prior expectactions of overall model sparsity and sparsity for the MRF graph.

\subsection{Dense Seemingly Unrelated Regression (dSUR)}  \label{sec:modeldSUR}

The HRR models in Section \ref{sec:modelHRR} assume residual independence between any two response variables because of the diagonal matrix $C$ in (\ref{diagC}). It is possible to estimate a full covariance matrix by specifying an inverse Wishart prior, i.e., $C \sim \mathcal{IW}(\nu, \tau \mathbb{I}_s)$. To avoid estimating the dense and large covariance matrix directly, \cite{Banterle2018} exploited a factorisation of the dense covariance matrix to transform the parameter space $(\nu,\tau)$ of the inverse Wishart distribution to space $\{\sigma_k^2,\rho_{kl}|\sigma_k^2 : k=1,\cdots,s; l<k\}$, with priors
\begin{align*} \tag{7}\label{IWdecomposition}
\begin{split}
\sigma_k^2 &\sim \mathcal{IG}amma\left(\frac{\nu-s+2k-1}{2}, \frac{\tau}{2}\right), \\
\rho_{kl} | \sigma_k^2 &\sim \mathcal{N}\left(0, \frac{\sigma_k^2}{\tau}\right).
\end{split}
\end{align*}
Here, we assume that $\tau \sim \mathcal{G}amma (a_{\tau}, b_{\tau})$. Thus, model (\ref{orginEq}) is rewritten as
\begin{align*}\tag{8}\label{reparameterisation} 
\begin{split}
\bm{y}_k &= \mathbf{X}\bm{\beta}_k + \sum_{l<k}\bm{u}_l\rho_{kl} + \bm{\epsilon}_k, \ \ k=1,\cdots,s, \\
\bm{\epsilon}_k &\sim \mathcal{N}(\mathbf{0},\ \sigma_k^2\mathbb{I}_n),
\end{split}
\end{align*}
where $\bm{u}_l = \bm{y}_l - \mathbf{X}\bm{\beta}_l$ and $\bm{\beta}_l$ is the $l$th column of $\bm{B}$, so again the likelihood is factorised across responses.

Similarly to the HRR model, employing either the simple independence prior (\ref{BernoulliPrior}), the hotspot prior (\ref{hotspotPrior}) or the MRF prior (\ref{MRFprior}) for the indicator matrix $\bm{\Gamma}$ results in different sparsity specifications for the regression coefficients in the dSUR model.
The marginal likelihood integrating out $\bm{B}$ is no longer available for this model, so joint sampling of $\bm{B}$, $\bm{\Gamma}$ and $C$ is required. However, the reparameterisation of the model (\ref{reparameterisation}) enables fast computation using the MCMC algorithm.

\subsection{Sparse Seemingly Unrelated Regression (SSUR)} \label{sec:modelSSUR}

Another approach to model the covariance matrix $C$ is to specify a hyper-inverse Wishart prior, which means the multiple response variables have an underlying graph $\mathcal{G}$ encoding the conditional dependence structure between responses. In this setup, a sparse graph corresponds to a sparse precision matrix $C^{-1}$. From a computational point of view, it is infeasible to specify a hyper-inverse Wishart prior directly on $C^{-1}$ in high dimensions \citep{Carvalho2007, Jones2005, Uhler2018, Deshpande2019}. 
However, \cite{Banterle2018} used a transformation of $C$ to factorise the likelihood as in equation (\ref{reparameterisation}).
The hyper-inverse Wishart distribution, i.e., $C \sim \mathcal{HIW}_{\mathcal{G}}(\nu, \tau \mathbb{I}_s)$, becomes in the transformed
variables
\begin{align*} \tag{10}
\begin{split}
\sigma_{qt}^2 &\sim \mathcal{IG}amma\left(\frac{\nu-s+t+|S_q|}{2}, \frac{\tau}{2}\right), \ \ q=1,\cdots,Q,t=1,\cdots,|R_q| \\
\bm{\rho}_{qt} |\sigma_{qt}^2 &\sim \mathcal{N}\left(\mathbf 0, \frac{\sigma_{qt}^2}{\tau}\mathbbm{I}_{t-1}\right)
\end{split}
\end{align*}
where $Q$ is the number of prime components in the decomposable graph $\mathcal{G}$, and $S_q$ and $R_q$ are the separators and residual components of $\mathcal{G}$, respectively. $|S_q|$ and $|R_q|$ denote the number of variables in these components. 
For more technical details, please refer to \cite{Banterle2018}.

As prior for the graph we use a Bernoulli prior with probility $\eta$ on each edge $E_{kk'}$ of the graph as in
\begin{align*} \tag{11}
\begin{split}
\mathbbm{P} (&E_{kk'} \in \mathcal{G}) = \eta, \\
\eta &\sim \mathcal{B}eta (a_{\eta}, b_{\eta}). 
\end{split}
\end{align*}
The three priors on $\bm{\beta}_{\bm{\gamma}}$, i.e., independence (\ref{BernoulliPrior}), hotspot (\ref{hotspotPrior}) and MRF (\ref{MRFprior}) priors can be used in the SSUR model.

\subsection{MCMC sampler and posterior inference} \label{sec:modelOutput}

To sample from the posterior distribution, we use 
the Evolutionary Stochastic Search algorithm \citep{Bottolo2010,Bottolo2011,Lewin2016}, which uses a particular form of Evolutionary Monte Carlo (EMC) introduced by \cite{Liang2000}. Multiple tempered Markov Chains are run in parallel and both exchange and crossover moves are allowed between the chains to improve mixing between potentially different modes in the posterior. Note that we run multiple tempered chains at the same temperature instead of a ladder of different temperatures as was proposed in the original implementations of the (H)ESS sampler in \citep{Bottolo2010,Bottolo2011,Lewin2016}. The temperature is adapted during the burn-in phase of the MCMC sampling.

The main chain samples from the un-tempered posterior distribution, which is used for all inference.
For each response variable, we use a Gibbs sampler to update the regression coefficients vector, $\bm{\beta}_{k}$ ($k=1,\cdots,s$), based on the conditional posterior corresponding to the specific model selected among the models presented in Section \ref{sec:modeldSUR}-\ref{sec:modelSSUR}. After $L$ MCMC iterations, we obtain $\bm{B}^{(1)},\cdots,\bm{B}^{(L)}$ and the estimate of the posterior mean is

$$\hat{\bm{B}} = \frac{1}{L-b}\sum_{t=b+1}^{L}\bm{B}^{(t)},$$

where $b$ is the number of burn-in iterations. Posterior full conditionals are also available to update $\sigma_{k}^2$ ($k=1,\cdots,s$) and $\rho_{kl} | \sigma_k^2 $ ($k=1,\cdots,s$ , $l<k$) for 
the dSUR and SSUR models. In the HRR models in Section \ref{sec:modelHRR}, the regression coefficients and residual covariances have been integrated out and therefore the MCMC output cannot be used directly for posterior inference of these parameters. However, for $\bm{B}$, the posterior distribution conditional on $\bm{\Gamma}$ can be derived analytically for the HRR models and this is the output for $\bm{B}$ that is provided in the \pkg{BayesSUR} package for HRR models.

At MCMC iteration $t$ we also update each binary latent vector $\bm{\gamma}_k$ ($k=1,\cdots,s$) via a Metropolis-Hastings sampler, jointly proposing an update for the corresponding $\bm{\beta}_{k}$. After $L$ iterations, using the binary matrices $\bm{\Gamma}^{(1)},\cdots,\bm{\Gamma}^{(L)}$, the marginal posterior inclusion probabilities (mPIP) of the predictors are estimated by

$$\hat{\bm{\Gamma}} = \frac{1}{L-b}\sum_{t=b+1}^{L}\bm{\Gamma}^{(t)}.$$

In the SSUR models, another important parameter is $\mathcal{G}$ in the hyper-inverse Wishart prior for the covariance matrix $C$. It is updated by the junction tree sampler \citep{Green2013,Banterle2018} jointly with the corresponding proposal for  $\bm{\sigma}_{k}^2,\bm{\rho}_{k},$ ($k=1,\cdots,s$). At each MCMC iteration we then extract the adjacency matrix $\mathcal{G}^{(t)}$ ($t=1,\cdots,L$), from which we derive posterior mean estimators of the edge inclusion probabilities as
$$\hat{\mathcal{G}} = \frac{1}{L-b}\sum_{t=b+1}^{L}\mathcal{G}^{(t)}.$$ Note that even though \emph{a priori} the graph $\mathcal{G}$ is decomposable, the posterior mean estimate $\hat{\mathcal{G}}$ can be outside the space of decomposable models (see \cite{Banterle2018}).

The hyper-parameter $\tau$ in the inverse Wishart prior or hyper-inverse Wishart prior is updated by a random walk Metropolis-Hastings sampler. The hyper-parameter $\eta$ and the variance $w$ in the spike-and-slab prior are sampled from their posterior conditional. For details see \cite{Banterle2018}.

\clearpage
\section{The R package BayesSUR} \label{sec:package}

The package \pkg{BayesSUR} is available from the Comprehensive \proglang{R} Archive Network (CRAN) at \url{http://CRAN.R-project.org/package=BayesSUR} and on GitHub \url{https://github.com/mbant/BayesSUR}. This article refers to version 1.2-4.

The main function is \code{BayesSUR()}, which has various arguments that can be used to specify the models introduced in Section \ref{sec:models}, by setting the priors for the covariance matrix $C$ and the binary latent indicator matrix $\bm{\Gamma}$. In addition, MCMC parameters (\code{nIter}, \code{burnin}, \code{nChains}) can also be defined. The following syntax example introduces the most important function arguments, which are further explained below. The full list of all arguments in function \code{BayesSUR()} is given in Table \ref{tab:arguments}. 

\begin{Schunk}
\begin{Sinput}
R> BayesSUR(data, Y, X, X_0, covariancePrior, gammaPrior,
+           nIter, burnin, nChains, ...)
\end{Sinput}
\end{Schunk}

The data can be provided as a large combined numeric matrix $[\mathbf{Y},\mathbf{X},\mathbf{X}_0]$ of dimension $n \times (s+p)$ via the argument \code{data}; in that case the arguments \code{Y}, \code{X} and \code{X\_0} need to contain the dimensions of the individual response variables $\mathbf{Y}$, predictors under selection $\mathbf{X}$ and fixed predictors $\mathbf{X}_0$ (i.e., mandatory predictors that will always be included in the model). Alternatively, it is also possible to supply $\mathbf{X}_0$, $\mathbf{X}$ and $\mathbf{Y}$ directly as numeric matrices via the arguments \code{X\_0}, \code{X} and \code{Y}. In that case, argument \code{data} needs to be \code{NULL}, which is the default. 

The arguments \code{covariancePrior} and \code{gammaPrior} specify the different models introduced in Section \ref{sec:models}. When using the Markov random field prior (\ref{MRFprior}) for the latent binary vector $\bm{\gamma}$, an additional argument \code{mrfG} is needed to assign the edge potentials; this can either be specified as a numeric matrix or as a file directory path leading to a text file with the corresponding information. For example, the HRR model with independent hierarchical prior in Section \ref{sec:modelHRRind} is specified by \code{(covariancePrior = "IG", gammaPrior = "hierarchical")}, the dSUR model with hotspot prior in Section \ref{sec:modeldSUR} by \code{(covariancePrior = "IW", gammaPrior = "hotspot")} and the SSUR model with MRF prior in Section \ref{sec:modelSSUR} for example by \\\code{(covariancePrior = "HIW", gammaPrior = "MRF", mrfG = "/mrfFile.txt")}.

The MCMC parameter arguments \code{nIter}, \code{burnin} and \code{nChains} indicate the total number of MCMC iterations, the number of iterations in the burn-in period and the number of parallel tempered chains in the evolutionary stochastic search MCMC algorithm, respectively. See Section \ref{sec:modelOutput} and e.g., \cite{Bottolo2010} for more details on the ESS algorithm.

\begin{table}[tph]
\centering
\begin{tabularx}{\linewidth}{l L } 
\hline
Argument & Description \\ [0.5ex] 
\hline
\code{data} &  Combined numeric data matrix $[\mathbf{Y},\mathbf{X}]$ or $[\mathbf{Y},\mathbf{X},\mathbf{X}_0]$. Default is \code{NULL}.\\
\code{Y} &  Numeric matrix or indices with respect to the argument \code{data} for the reponses.\\
\code{X} &  Numeric matrix or indices with respect to the argument \code{data} for the predictors.\\
\code{X\_0} & Numeric matrix or indices with respect to the argument \code{data} for predictors forced to be included (i.e., that are not part of variable selection procedure). Default is \code{NULL}. \\
\code{outFilePath} & Directory path where the output files are saved.  Default is the current working directory. \\
\code{covariancePrior} & Prior for the covariance matrix; "\code{IG}": independent inverse gamma prior, "\code{IW}": inverse Wishart prior, "\code{HIW}": hyper-inverse Wishart prior (default).\\
\code{gammaPrior} &  Prior for the binary latent variable $\bm\Gamma$; "\code{hierarchical}": Bernoulli prior, "\code{hotspot}": hotspot prior (default), "\code{MRF}": Markov random field prior. \\
\code{mrfG} & A numeric matrix or a path to the file containing the edge list of the G matrix for the MRF prior on $\bm\Gamma$. Default is \code{NULL}. \\
\code{nIter} & Total number of MCMC iterations.\\
\code{burnin} & Number of iterations in the burn-in period.\\
\code{nChains} & Number of parallel chains in the evolutionary stochastic search MCMC algorithm.\\
\code{gammaSampler} & Local move sampler for the binary latent variable $\bm\Gamma$ , either (default) "\code{bandit}" for a Thompson-sampling inspired sampler or "\code{MC3}" for the usual $MC^3$ sampler. \\
\code{gammaInit} & $\mathbf{\Gamma}$ initialisation to either all zeros ("\code{0}"), all ones ("\code{1}"), MLE-informed ("\code{MLE}") or (default) randomly ("\code{R}"). \\
\code{hyperpar} & A list of named prior hyperparameters to use instead of the default values, including \code{a\_w}, \code{b\_w}, \code{a\_sigma}, \code{b\_sigma}, \code{a\_omega}, \code{b\_omega}, \code{a\_o}, \code{b\_o}, \code{a\_pi}, \code{b\_pi}, \code{nu}, \code{a\_tau}, \code{b\_tau}, {a\_eta} and \code{b\_eta}. They correspond to $w\sim \mathcal{IG}amma$(\code{a\_w}, \code{b\_w}), $\sigma_k^2\sim \mathcal{IG}amma$(\code{a\_sigma}, \code{b\_sigma}), $\omega_j\sim \mathcal{B}eta$(\code{a\_omega}, \code{b\_omega}), $o_k\sim \mathcal{B}eta$(\code{a\_o}, \code{b\_o}), $\pi_j\sim \mathcal{G}amma$(\code{a\_pi}, \code{b\_pi}), $\nu$=\code{nu}, $\tau\sim \mathcal{G}amma$(\code{a\_tau}, \code{b\_tau}), $\eta \sim \mathcal{B}eta$(\code{a\_eta}, \code{b\_eta}). For default values see \code{help(BayesSUR)}. \\
\code{maxThreads} & Maximum threads used for parallelisation. Default is 1.\\
\code{output\_}* & Allow (\code{TRUE}) or suppress (\code{FALSE}) the outut for *; possible outputs are $\bm\Gamma$ , $\mathcal{G}$, $\bm{B}$, $\bm{\sigma}$, $\bm{\pi}$, tail (hotspot tail probability, see \cite{Bottolo2010}) or model\_size. Default is all \code{TRUE}.\\
\code{tmpFolder} & The path to a temporary folder where intermediate data files are stored (will be erased at the end of the MCMC sampling). Defaults to local \code{tmpFolder}. \\
\hline
\end{tabularx}
\caption{Overview of the arguments in the main function \code{BayesSUR()}}
\label{tab:arguments}
\end{table}

\begin{table}
\centering
\begin{tabularx}{\linewidth}{l L } 
\hline
Function & Description \\ [0.5ex] 
\hline
\code{BayesSUR()} & Main function of the package. Fits any of the models introduced in Section \ref{sec:models}. Returns an object of \code{S3} class \code{BayesSUR}, which is a list which includes the input parameters (input) and directory paths of output text files (output), as well as the run status and function call.\\

\code{print()} & Print a short summary the fitted model generated by \code{BayesSUR()}, which is an object of class \code{BayesSUR}. \\
\code{summary()} & Summarise the fitted model generated by \code{BayesSUR()}, which is an object of class \code{BayesSUR}. \\
\code{coef()} & Extract the posterior mean of the coefficients of a \code{BayesSUR} class object. \\
\code{fitted()} & Return the fitted response values that correspond to the posterior mean of the coefficients matrix of a \code{BayesSUR} class object. \\
\code{predict()} & Predict responses corresponding to the posterior mean of the coefficients, return posterior mean of coefficients or indices of nonzero coefficients of a \code{BayesSUR} class object. \\
\code{plot()} & Main plot function to be called by the user. Creates a selection of plots for a \code{BayesSUR} class object by calling one or several of the specific plot functions below as specified by the combination of the two arguments \code{estimator} and \code{type}. \\

\code{elpd()} & Measure the prediction accuracy by the expected log pointwise predictive density (elpd). The out-of-sample predictive fit can either be estimated by Bayesian leave-one-out cross-validation (LOO) or by widely applicable information criterion (WAIC) \citep{Vehtari2017}. See Appendix for details. \\
\code{getEstimator()} & Extract the posterior mean of the parameters (i.e., $\bm{B}$, $\mathbf{\Gamma}$ and $\mathcal{G}$) of a \code{BayesSUR} class object. Also, the log-likelihood of $\mathbf{\Gamma}$, model size and $\mathcal{G}$ can be extracted for the MCMC diagnostics. \\

\code{plotEstimator()} & Plot the estimated relationships between response variables and estimated coefficients of a \code{BayesSUR} class object with argument \code{estimator=c("beta","gamma","Gy")}.\\
\code{plotGraph()} & Plot the estimated graph for multiple response variables from a \code{BayesSUR} class object with argument \code{estimator="Gy"}.  \\
\code{plotNetwork()} & Plot the network representation of the associations between responses and predictors, based on the estimated $\hat{\bm\Gamma}$ matrix of a \code{BayesSUR} class object with argument \code{estimator=c("gamma","Gy")}. \\
\code{plotManhattan()} & Plot Manhattan-like plots for marginal posterior inclusion probabilities (mPIP) and numbers of responses of association for predictors of a \code{BayesSUR} class object with argument \code{estimator="gamma"}.\\
\code{plotMCMCdiag()} & Show trace plots and diagnostic density plots of a \code{BayesSUR} class object with argument \code{estimator="logP"}. \\
\code{plotCPO()} & Plot the conditional predictive ordinate (CPO) for each individual of a fitted model generated by \code{BayesSUR} with argument \code{estimator="CPO"}. CPO is used to identify potential outliers \citep{Gelfand1996}. \\

\hline
\end{tabularx}
\caption{Overview of the functions in package \pkg{BayesSUR}.}
\label{tab:functions}
\end{table}

The main function \code{BayesSUR()} is used to fit the model. It returns an object of \code{S3} class \code{BayesSUR} in a list format, which includes the input parameters and directory paths of output text files, so that other functions can retrieve the MCMC output from the output files, load them into \proglang{R} and further process the output for posterior inference of the model output.

In particular, a \code{summary()} function has been provided for \code{BayesSUR} class objects, which is used to summarise the output produced by \code{BayesSUR()}. For this purpose, a number of predictors are selected into the model by thresholding the posterior means of the latent indicator variables. By default, the threshold is $0.5$, i.e., variable $j$ is selected into the model for response $k$ if $\hat{\gamma}_{jk}>0.5$. The \code{summary()} function also outputs the quantiles of the conditional predictive ordinates (CPO, \cite{Gelfand1996}), top predictors with respect to average marginal posterior inclusion probabilities (mPIP) across all response variables and top response variables with respect to average mPIP across all predictors, expected log pointwise predictive density (i.e., \code{elpd.LOO} and \code{elpd.WAIC}, \cite{Vehtari2017}), model specification parameters, MCMC running parameters and hyperparameters.

To use a specific estimator, the function \code{getEstimator()} is convenient to extract point estimates of the coefficients matrix $\hat{\bm{B}}$, latent indicator variable matrix $\hat{\bm\Gamma}$ or learned structure $\hat{\mathcal{G}}$ from the directory path of the model object. 
All point estimates are posterior means, thus $\hat{\gamma}_{jk}$ is the marginal posterior inclusion probability for variable $j$ to be selected 
in the regression for response $k$, and $\hat{\mathcal{G}}_{kl}$ is the marginal posterior edge inclusion probability between responses $k$ and $l$,
i.e., the marginal posterior probability of conditional dependence between $k$ and $l$.
The regression coefficient estimates $\hat{\bm{B}}$ can be the marginal posterior means over all models, independently of $\hat{\mathbf{\Gamma}}$ (with default argument \code{beta.type = "marginal"}). Then, $\hat{\beta}_{jk}$ represents the shrunken estimate of the effect of variable $j$ in the regression for response $k$. Alternatively, $\hat{\beta}_{jk}$ can be the posterior mean conditional on $\gamma_{jk} = 1$ with argument \code{beta.type = "conditional"}. If \code{beta.type="conditional"} and \code{Pmax = 0.5} are chosen, then these conditional $\hat{\beta}_{jk}$ estimates correspond to the coefficients in a median probability model \citep{Barbieri2004}.

In addition, the generic \code{S3} methods \code{coef()}, \code{predict()}, and \code{fitted()} can be used to extract regression coefficients, predicted responses, or indices of nonzero coefficients, all corresponding to the posterior mean estimates of an \code{BayesSUR} object.

The main function for creating plots of a fitted BayesSUR model, is the generic \code{S3} method \code{plot()}. It creates a selection of the above plots, which the user can specify via the \code{estimator} and \code{type} arguments. If both arguments are set to \code{NULL} (default), then all available plots are shown in an interactive manner. The main \code{plot()} function uses the following specific plot functions internally. These can also be called directly by the user.
The function \code{plotEstimator()} visualises the three estimators. To show the relationship of multiple response variables with each other, the function \code{plotGraph()} prints the structure graph based on $\hat{\mathcal{G}}$. Furthermore, the structure relations between multiple response variables and predictors can be shown via function \code{plotNetwork()}. The marginal posterior probabilities of individual predictors are illustrated via the \code{plotManhattan()} function, which also shows the number of associated response variables of each predictor. 

Model fit can be investigated with \code{elpd()} and \code{plotCPO()}. \code{elpd()} estimates the expected log pointwise predictive density \citep{Vehtari2017} to assess out-of-sample prediction accuracy. \code{plotCPO()} plots the conditional predictive ordinate for each individual, i.e., the leave-one-out cross-validation predictive density. CPOs are useful for identifying potential outliers \citep{Gelfand1996}. To check convergence of the MCMC sampler, function \code{plotMCMCdiag()} prints traceplots and density plots for moving windows over the MCMC chains. 

Table \ref{tab:functions} lists all functions. \pkg{BayesSUR} uses the \pkg{Rcpp} \citep{Edelbuettel_2011} and \pkg{RcppArmadillo} \citep{Edelbuettel_2014} \proglang{R} packages to integrate \proglang{C++} code with \proglang{R}. The \pkg{igraph} package \citep{Csardi_2006} was used for constructing the graph plots. 

\section[Quick start with a simple example]{Quick start with a simple example} \label{sec:tutorial}

In the following example, we illustrate a simple simulation study where we run two models: the default model choice, which is an SSUR model with the hotspot prior, and in addition an SSUR model with the MRF prior. The purpose of the latter is to illustrate how we can construct an MRF prior graph. We simulate a dataset $\mathbf{X}$ with dimensions $n \times p = 10 \times 15$, i.e., 10 observations and 15 input variables, a sparse coefficients matrix $\bm{B}$ with dimension $p \times s = 15 \times 3$, which creates associations between the input variables and $s=3$ response variables, and random noise $\mathbf{E}$. The response matrix is generated by the linear model $\mathbf{Y} = \mathbf{X}\bm{B} +\mathbf{E}$.

\begin{Schunk}
\begin{Sinput}
R> set.seed(82193)
R> n <- 10; s <- 3; p <- 15
R> X <- matrix(rnorm(n * p, 2, 1), nrow = n)
R> B <- matrix(c(0, 0, 1,
+                1, 1, 0,
+                1, 1, 0,
+                0, 0, 1,
+                rep(0, s * p - 12)), 
+              nrow = p, byrow = TRUE)
R> E <- matrix(rnorm(n * s, 0, 0.2), nrow = n)
R> Y <- X %*% B + E
\end{Sinput}
\end{Schunk}

Note that $\bm{B}$ is sparse and only the first three input variables have non-zero coefficients:
\begin{Schunk}
\begin{Sinput}
R> print(B)
\end{Sinput}
\begin{Soutput}
      [,1] [,2] [,3]
 [1,]    0    0    1
 [2,]    1    1    0
 [3,]    1    1    0
 [4,]    0    0    1
 [5,]    0    0    0
 [6,]    0    0    0
 [7,]    0    0    0
 [8,]    0    0    0
 [9,]    0    0    0
[10,]    0    0    0
[11,]    0    0    0
[12,]    0    0    0
[13,]    0    0    0
[14,]    0    0    0
[15,]    0    0    0
\end{Soutput}
\end{Schunk}

First, let's fit the default model. The default is to run two MCMC chains with 10000 iterations each, of which the first 5000 iterations are discarded as the burn-in period. The function \code{print()} returns a short summary of the results from the fitted model object, including the number of selected predictors by thresholding the marginal posterior inclusion probabilities (mPIP) at 0.5, and two measures of the model's prediction accuracy (i.e., \code{elpd.LOO} and \code{elpd.WAIC}).

\begin{Schunk}
\begin{Sinput}
R> library("BayesSUR")
R> library("tictoc")
R> tic("Time of model fitting")
R> set.seed(1294)
R> fit <- BayesSUR(Y = Y, X = X, outFilePath = "results/", output_CPO = TRUE)
\end{Sinput}
\end{Schunk}
\vspace{-8.8mm}
\begin{Schunk}
\begin{Sinput}
R> toc()
\end{Sinput}
\begin{Soutput}
Time of model fitting: 2.755 sec elapsed
\end{Soutput}
\begin{Sinput}
R> print(fit)
\end{Sinput}
\begin{Soutput}
Call:
 BayesSUR(Y = Y, X = X, outFilePath = "results/", ...)

Number of selected predictors (mPIP > 0.5): 6 of 3x15

Expected log pointwise predictive density (elpd):
 elpd.LOO = -40.84189,  elpd.WAIC = -41.12811
\end{Soutput}
\end{Schunk}

\begin{figure}[tph]
\centering
\includegraphics[height=6cm,keepaspectratio]{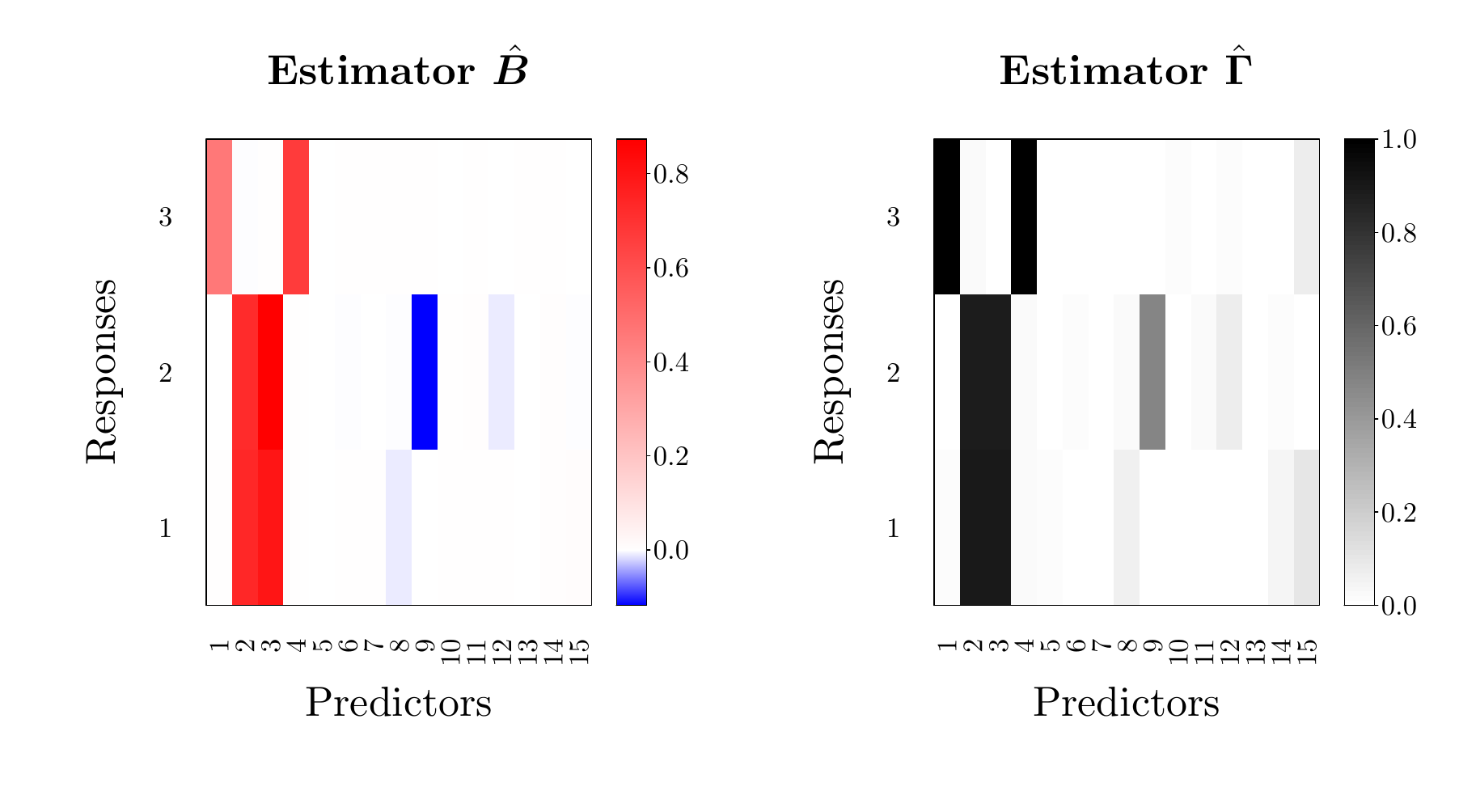}
\caption{The posterior mean estimates of the coefficients matrix $\hat{\bm{B}}$ and latent indicator matrix $\hat{\bm{\Gamma}}$ for the SSUR model with the hotspot prior plotted with \code{plot(fit, estimator = c("beta", "gamma"), type = "heatmap", ...)}.}
\label{TutorialParamEstimator1}
\end{figure}

The posterior means of the coefficients and latent indicator matrices are printed by the function \code{plot()} with arguments \code{estimator = c("beta", "gamma")} and \code{type = "heatmap"} (Figure \ref{TutorialParamEstimator1}). Note, that the argument \code{fig.tex = TRUE} produces PDF figures through LaTeX with the \code{tools::texi2pdf()} function, which creates authentic math formulas in the figure labels, but requires that the user has LaTeX installed. The argument \code{output} specifies the name of the PDF file.

\begin{Schunk}
\begin{Sinput}
R> plot(fit, estimator = c("beta", "gamma"), type = "heatmap", fig.tex = TRUE, 
+       output = "exampleEst", xlab = "Predictors", ylab = "Responses")
\end{Sinput}
\end{Schunk}

Before running the SSUR model with the MRF prior, we need to construct the edge potentials matrix $G$. If we assume (in accordance with the true matrix $\bm{B}$ in this simulation scenario) that the second and third predictors are related to the first two response variables, this implies that $\gamma_{21}$, $\gamma_{22}$, $\gamma_{31}$ and $\gamma_{32}$ are expected to be related and therefore we might want to encourage these variables to be selected together. In addition, we assume that we know that the 1st predictor is associated with the 3rd response variable, and therefore we encourage the selection of $\gamma_{13}$ as well. Since matrix $G$ represents prior relations of any two predictors corresponding to vec$\{\bm{\Gamma}\}$, it can be generated by the following code: 

\begin{Schunk}
\begin{Sinput}
R> G <- matrix(0, ncol = s * p, nrow = s * p)
R> combn1 <- combn(rep((1:2 - 1) * p, each = length(2:3)) 
+                  + rep(2:3, times = length(1:2)), 2)
R> combn2 <- combn(rep((3-1) * p, each = length(c(1,4))) 
+                + rep(c(1,4), times = length(3)), 2)
R> G[c(combn1[1,], combn2[1]), c(combn1[2,], combn2[2])] <- 1
\end{Sinput}
\end{Schunk}

\begin{figure}[tph]
\centering
\includegraphics[height=6cm,keepaspectratio]{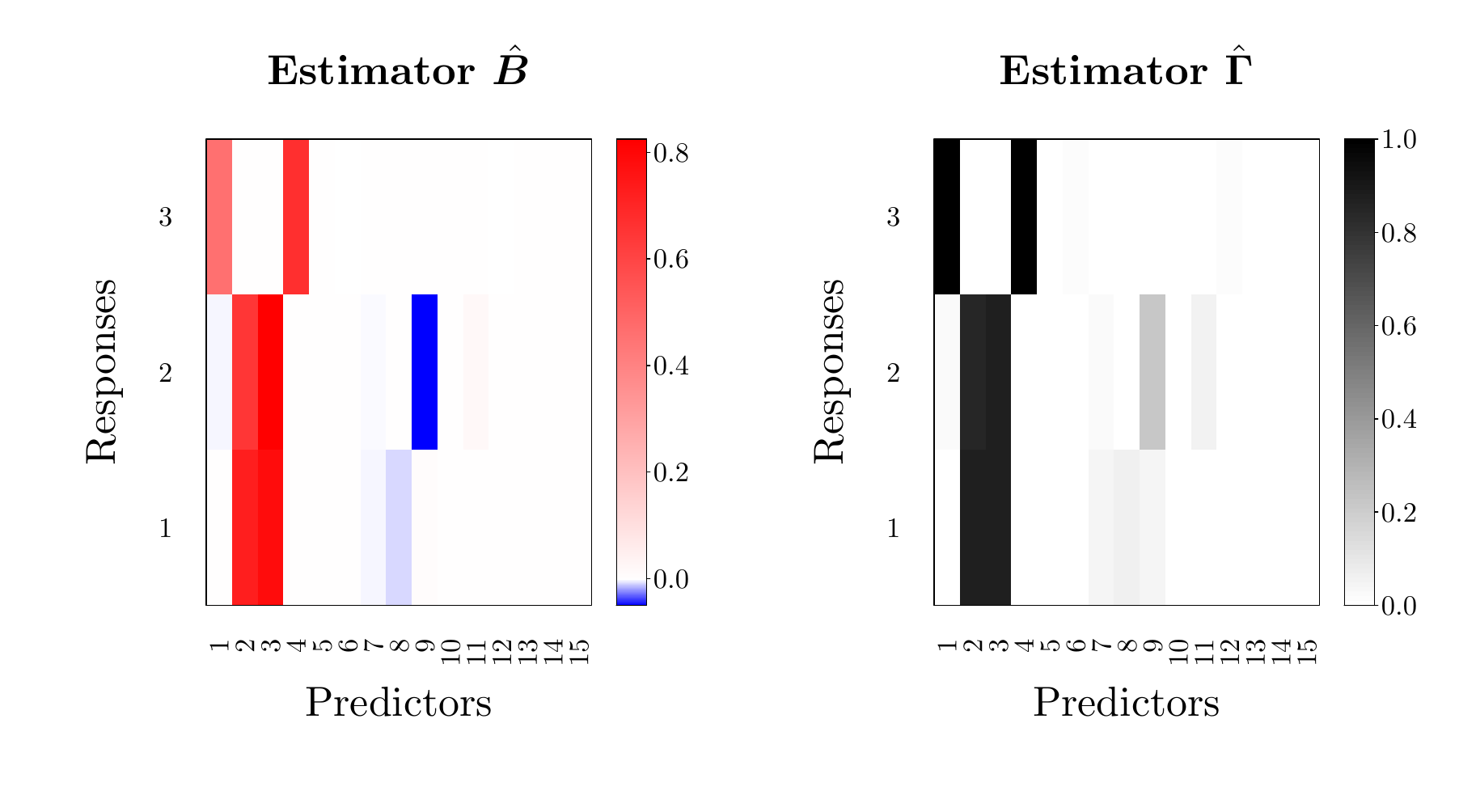}
\caption{The posterior mean estimates of the coefficients matrix $\hat{\bm{B}}$ and latent indicator matrix $\hat{\bm{\Gamma}}$ for the SSUR model with MRF prior plotted with \code{plot(fit, estimator = c("beta", "gamma"), type = "heatmap", ...)}.}
\label{TutorialParamEstimator2}
\end{figure}

Calling \code{BayesSUR()} with the argument \code{gammaPrior = "MRF"} will run the SSUR model with the MRF prior, and the argument \code{mrfG = G} imports the edge potentials for the MRF prior. The two hyper-parameters $d$ and $e$ for the MRF prior (\ref{MRFprior}) can be specified through the argument \code{hyperpar}; here we use the default values $d = -3$, $e = 0.03$. The posterior mean estimates for the coefficients matrix and latent indicator matrix are shown in Figure \ref{TutorialParamEstimator2}.
\begin{Schunk}
\begin{Sinput}
R> tic("Time of model fitting")
R> set.seed(5294)
R> fit <- BayesSUR(Y = Y, X = X, outFilePath = "results/",
+               gammaPrior = "MRF", mrfG = G)
\end{Sinput}
\end{Schunk}
\vspace{-8.8mm}
\begin{Schunk}
\begin{Sinput}
R> toc()
\end{Sinput}
\begin{Soutput}
Time of model fitting: 2.506 sec elapsed
\end{Soutput}
\end{Schunk}
\begin{Schunk}
\begin{Sinput}
R> plot(fit, estimator = c("beta", "gamma"), type = "heatmap", fig.tex = TRUE, 
+       output = "exampleEst2", xlab = "Predictors", ylab = "Responses")
\end{Sinput}
\end{Schunk}

\section{Two extended examples based on real data} \label{sec:examples}

In this section, we use a simulated eQTL dataset and real data from a pharmacogenomic database to illustrate the usage of the \pkg{BayesSUR} package. The first example is under the known true model and demonstrates the recovery performance of the models introduced in Section \ref{sec:models}. It also demonstrates a full data analysis step by step. The second example illustrates how to use potential relationships between multiple response variables and input predictors as the prior information in Bayesian SUR models and showcases how the resulting estimated graph structures can be visualised with functions provided in the package.

\subsection{Simulated eQTL data}

\begin{figure}[tph]
\centering
\includegraphics[height=6cm,keepaspectratio]{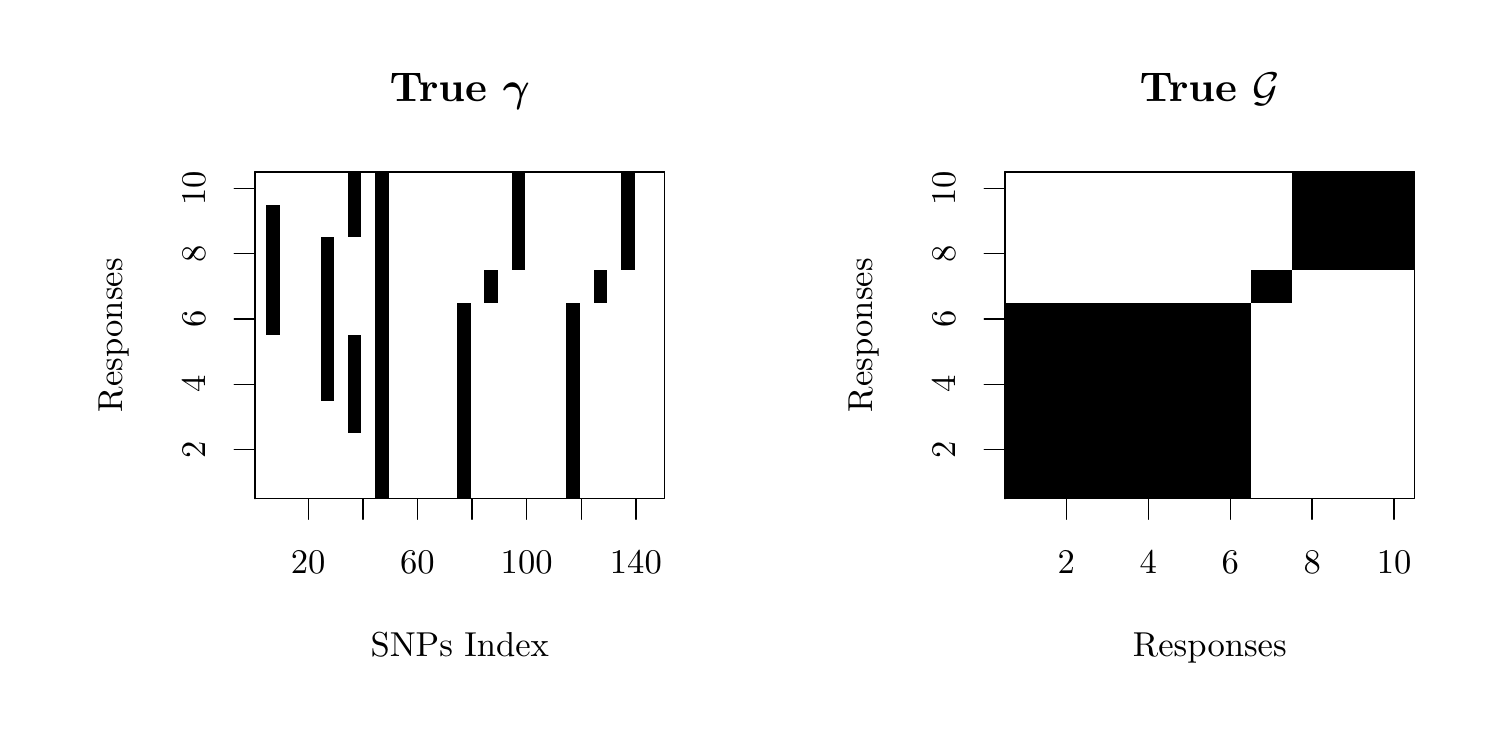}
\caption{True parameters of the simulated dataset \code{exampleEQTL}. The left panel is the designed sparse $\bm\Gamma$  and the right panel is the given true structure of responses represented by the decomposible graph $\mathcal{G}$. Black indicates a value 1 and white indicates 0.}
\label{ParamTrue}
\end{figure}

Similarly to \cite{Banterle2018}, we simulate single nucleotide polymorphism (SNP) data $\mathbf{X}$ by resampling from the \textbf{scrime} package \citep{Schwender2018}, with $p=150$ SNPs and $n=100$ subjects. To construct multiple response variables $\mathbf{Y}$ (with $s=10$) with structured correlation - which we imagine to represent gene expression measurements of genes that are potentially affected by the SNPs - we first fix a sparse latent indicator variable $\bm{\Gamma}$ and then design a decomposable graph for responses to build association patterns between multi-response variables and predictors. The nonzero coefficients are sampled from the normal distribution independently and the noise term from a multivariate normal distribution with the precision matrix sampled from $\mathcal{G}$-Wishart distribution $\mathcal{W}_{\mathcal{G}}(2,M)$ (\cite{Mohammadi2019}). Finally, the simulated gene expression data $\mathbf{Y}$ is then generated from the linear model \eqref{orginEq}. The concrete steps are as follows:

\begin{itemize}
	\item Simulate SNPs data $\mathbf{X}$ from the \proglang{scrime} package, dim$(\mathbf{X})=n \times p$.
	\item Design a decomposable graph $\mathcal{G}$ as the right panel of Figure {\ref{ParamTrue}}, dim$(\mathcal{G})=s \times s$.
	\item Design a sparse matrix $\bm\Gamma$  as the left panel of Figure {\ref{ParamTrue}}, dim$(\bm\Gamma)=p \times s$.
	\item Simulate $\beta_{jk} \sim \mathcal{N}(0,1)$, $j=1,\cdots,p$ and $k=1,\cdots,s$.
	\item Simulate $\tilde{u}_{ij} \sim \mathcal{N}(0,0.5^2)$, $i=1,\cdots,n$ and $j=1,\cdots,p$.
	\item Simulate $P\sim \mathcal{W}_{\mathcal{G}}(2,M)$ where diagonals of $M$ are 1 and off-diagonals are 0.9, dim$(P)=s \times s$.
	\item Use Cholesky decomposition $\text{chol}(P^{-1})$ to get $\mathbf{U} = \tilde{\mathbf{U}} \cdot \text{chol}(P^{-1})$. 
	\item Generate $\mathbf{Y} = (\mathbf{X}\bm{B})_{\mathbf{\Gamma}} + \mathbf{U}$.
\end{itemize}

The resulting average signal-to-noise ratio is 25. The \proglang{R} code for the simulation can be found through \code{help("exampleEQTL")}. 

\begin{Schunk}
\begin{Sinput}
R> data("exampleEQTL", package = "BayesSUR")
R> str(exampleEQTL)
\end{Sinput}
\begin{Soutput}
List of 4
 $ data     : num [1:100, 1:160] -0.185 -1.01 -2.102 -2.88 1.749 ...
  ..- attr(*, "dimnames")=List of 2
  .. ..$ : chr [1:100] "1" "2" "3" "4" ...
  .. ..$ : chr [1:160] "GEX1" "GEX2" "GEX3" "GEX4" ...
 $ blockList:List of 2
  ..$ : int [1:10] 1 2 3 4 5 6 7 8 9 10
  ..$ : num [1:150] 11 12 13 14 15 16 17 18 19 20 ...
 $ gamma    : num [1:150, 1:10] 0 0 0 0 0 0 0 0 0 0 ...
 $ Gy       : num [1:10, 1:10] 1 1 1 1 1 1 0 0 0 0 ...
  ..- attr(*, "dimnames")=List of 2
  .. ..$ : NULL
  .. ..$ : chr [1:10] "GEX1" "GEX2" "GEX3" "GEX4" ...
\end{Soutput}
\end{Schunk}
\begin{Schunk}
\begin{Sinput}
R> attach(exampleEQTL)
\end{Sinput}
\end{Schunk}

In the \pkg{BayesSUR} package, the data $\mathbf{Y}$ and $\mathbf{X}$ are provided as a numeric matrix in the first list component \code{data} of the example dataset \code{exampleEQTL}. Here the first 10 columns of \code{data} are the $\mathbf{Y}$ variables, and the last 150 columns are the $\mathbf{X}$ variables. The second component of \code{exampleEQTL} is \code{blockList} which specifies the indices of $\mathbf{Y}$ and $\mathbf{X}$ in \code{data}. The third component is the true latent indicator matrix $\mathbf{\Gamma}$ of regression coefficients. The fourth component is the true graph $\mathcal{G}$ between response variables. Throughout this section we attach the data set for more concise \proglang{R} code.

Figure \ref{ParamTrue} shows the true $\bm\Gamma$  and decomposible graph $\mathcal{G}$ used in the eQTL simulation scenario. The following code shows how to fit an SSUR model with hotspot prior for the indicator variables $\bm\Gamma$  and the sparsity-inducing hyper-inverse Wishart prior for the covariance using the main function \code{BayesSUR()}. 

\begin{Schunk}
\begin{Sinput}
R> set.seed(28173)
R> tic("Time of model fitting")
R> fit <- BayesSUR(data = data, Y = blockList[[1]], X = blockList[[2]], 
+                outFilePath = "results/", nIter = 200000, nChains = 3, 
+                burnin = 100000, covariancePrior = "HIW", 
+                gammaPrior = "hotspot")
\end{Sinput}
\end{Schunk}
\vspace{-8.8mm}
\begin{Schunk}
\begin{Sinput}
R> toc()
\end{Sinput}
\end{Schunk}
\code{Time of model fitting: 1159.871 sec elapsed}\\

\begin{figure}[tph]
\centering
\includegraphics[height=9cm,keepaspectratio]{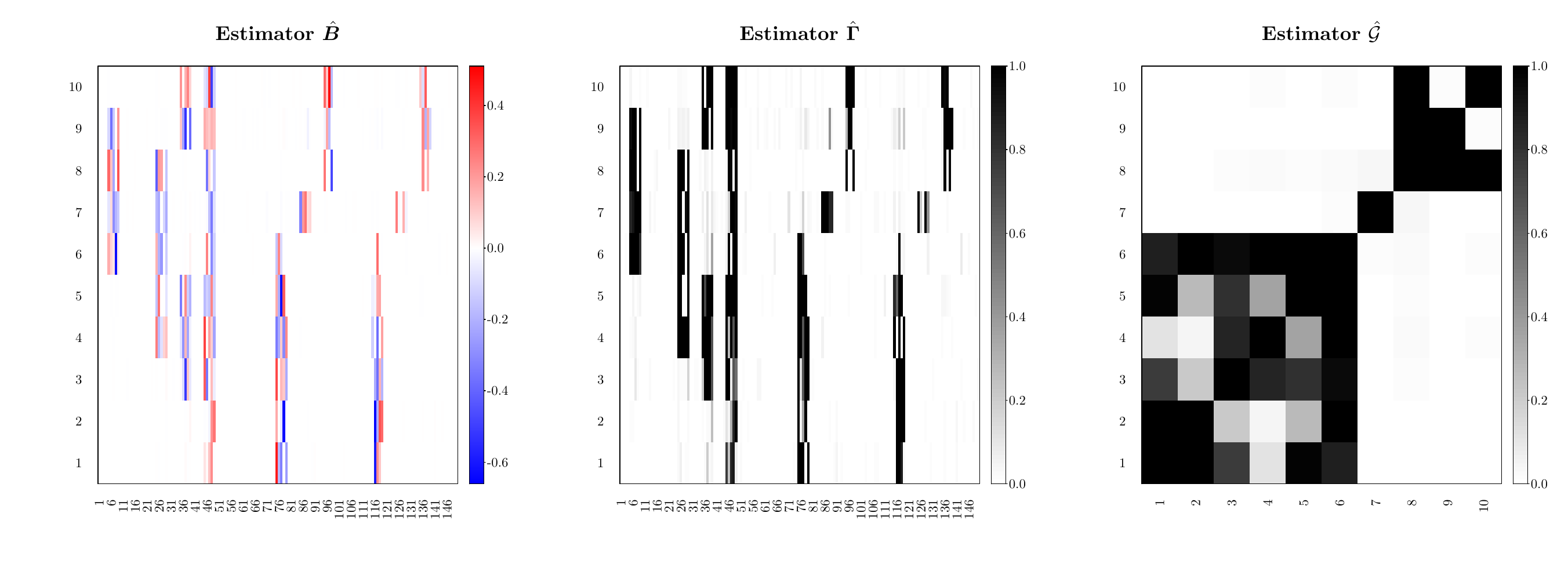}
\caption{The estimated coefficients matrix $\hat{\bm{B}}$, latent indicator variable matrix $\hat{\bm{\Gamma}}$ and learned structure $\hat{\mathcal{G}}$ of the SSUR model with hotspot prior and sparse covariance prior by \code{plot()}.}
\label{ParamEstimator}
\end{figure}

\begin{figure}[tph]
\centering
\includegraphics[height=6cm,keepaspectratio]{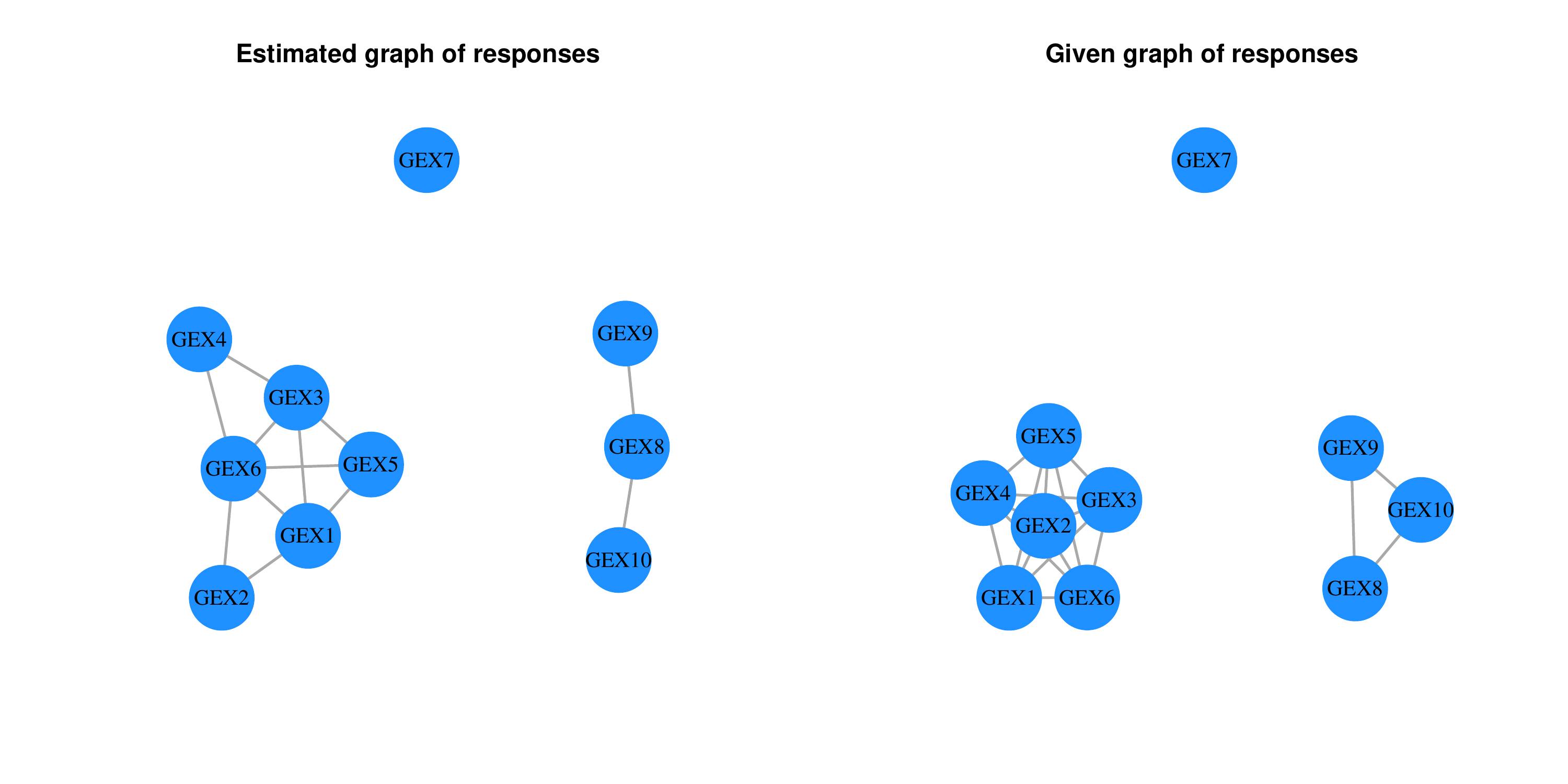}
\caption{The estimated structure of the ten response variables is visualised by \code{plot(fit, estimator = "Gy", type="graph")} with $\hat{\mathcal{G}}$ thresholded at 0.5 (left). The true structure is shown with \code{plotGraph(Gy)}, where \code{Gy} is the true adjacency matrix (right).}
\label{ResponseGraph}
\end{figure}

\begin{figure}[tph]
\centering
\includegraphics[height=12cm,keepaspectratio]{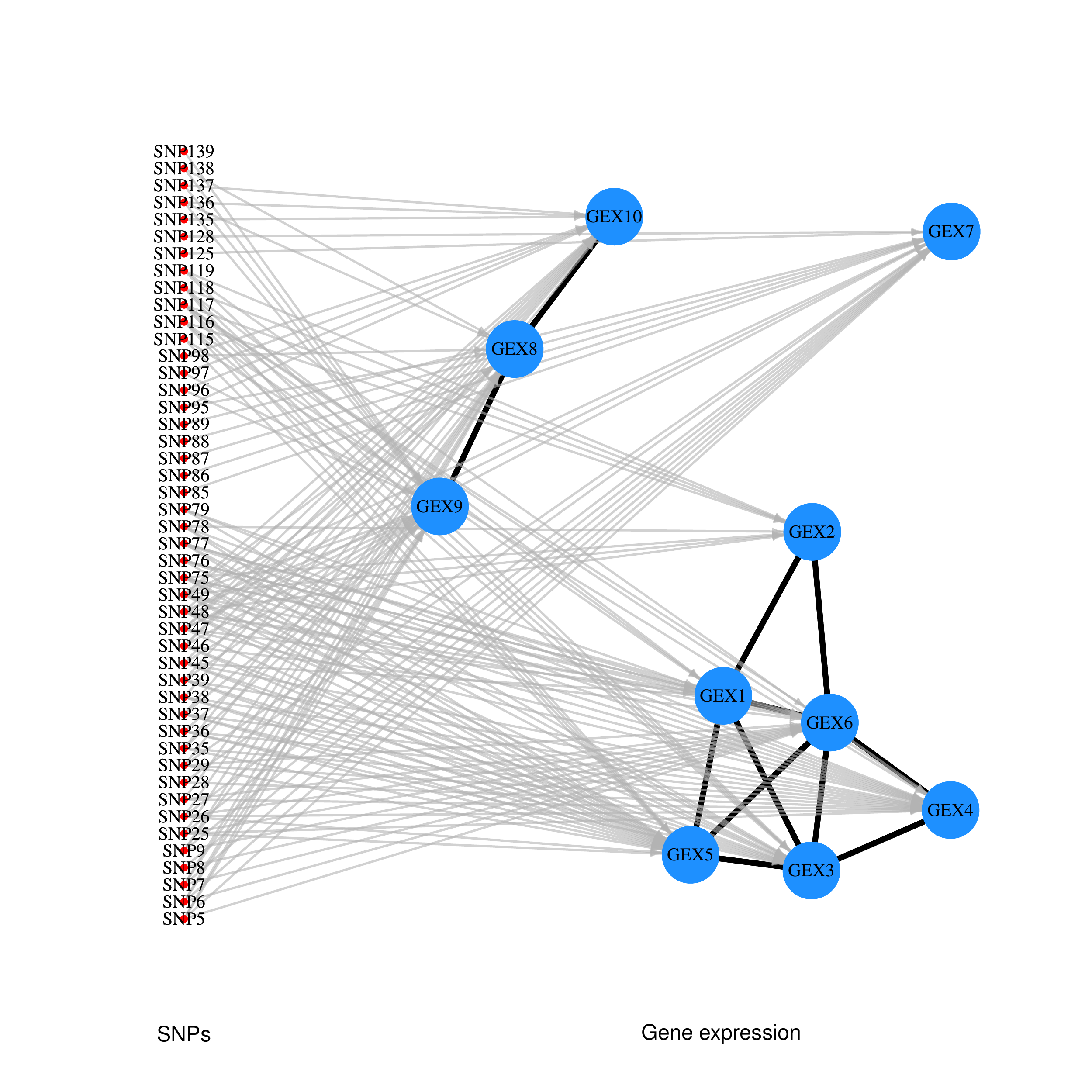}
\caption{Network representation between the ten gene expression variables and 150 SNP variables by \code{plot(fit, estimator = c("gamma","Gy"), type = "network", ...)}. The connections between gene expression variables are based on $\hat{\mathcal{G}}$ thresholded at 0.5, and the connections between the gene expression variables and SNPs are based on $\hat{\bm\Gamma}$ thresholded at 0.5.}
\label{Network}
\end{figure}

\begin{figure}[tph]
\centering
\includegraphics[height=10cm,keepaspectratio]{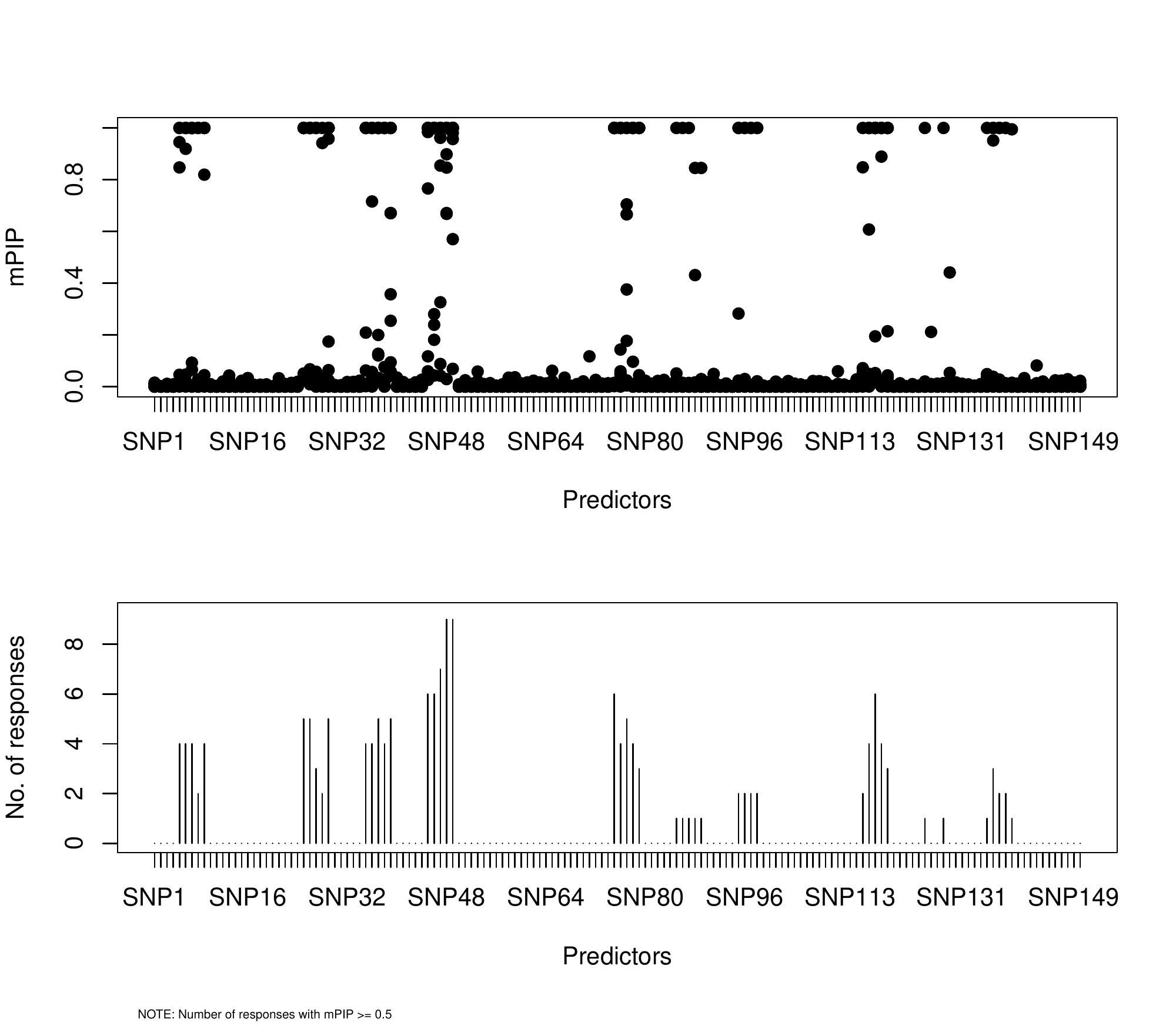}
\caption{Manhattan-like plots by \code{plot(fit, estimator = "gamma", type = "Manhattan")}. The top panel shows the marginal posterior inclusion probabilities (mPIP) of each SNP, and the bottom panel shows the number of gene expression response variables associated with each SNP. The number of responses are based on $\hat{\bm\Gamma}$ thresholded at 0.5.}
\label{Manhattan}
\end{figure}

Figure \ref{ParamEstimator} summarises the posterior inference results by plots for $\hat{\bm{B}}$, $\hat{\bm\Gamma}$ and $\hat{\mathcal{G}}$ created with the function \code{plot()} with arguments \code{estimator = c("beta","gamma","Gy")} and \code{type = "heatmap"}. When comparing with Figure \ref{ParamTrue}, we see that this SSUR model has good recovery of the true latent indicator matrix $\bm\Gamma$ and of the structure of the responses as represented by $\mathcal{G}$. The function \code{plot()} can also visualise the estimated structure of the ten gene expression variables as shown in the right panel of Figure \ref{ResponseGraph} with arguments \code{estimator = "Gy"} and \code{type = "graph"}. For comparison, the true structure is shown in the left panel (created by function \code{plotGraph()}). When we threshold the posterior selection probability estimates for $\mathcal{G}$ and for $\bm\Gamma$ at 0.5, the resulting full network between the ten gene expression variables and 150 SNPs is displayed in Figure \ref{Network}. Furthermore, the Manhattan-like plots in Figure \ref{Manhattan} show both, the marginal posterior inclusion probabilities (mPIP) of the SNP variables (top panel) and the number of gene expression response variables associated with each SNP (bottom panel).

\begin{Schunk}
\begin{Sinput}
R> plot(fit, estimator = c("beta","gamma","Gy"), type = "heatmap",
+       fig.tex = TRUE)
\end{Sinput}
\end{Schunk}

\begin{Schunk}
\begin{Sinput}
R> layout(matrix(1:2, ncol=2))
R> plot(fit, estimator = "Gy", type="graph")
R> plotGraph(Gy)
\end{Sinput}
\end{Schunk}

\begin{Schunk}
\begin{Sinput}
R> plot(fit, estimator = c("gamma","Gy"), type = "network", 
+       name.predictors = "SNPs", name.responses = "Gene expression")
\end{Sinput}
\end{Schunk}

\begin{Schunk}
\begin{Sinput}
R> plot(fit, estimator = "gamma", type = "Manhattan")
\end{Sinput}
\end{Schunk}

\begin{figure}[tph]
\centering
\includegraphics[height=12cm,keepaspectratio]{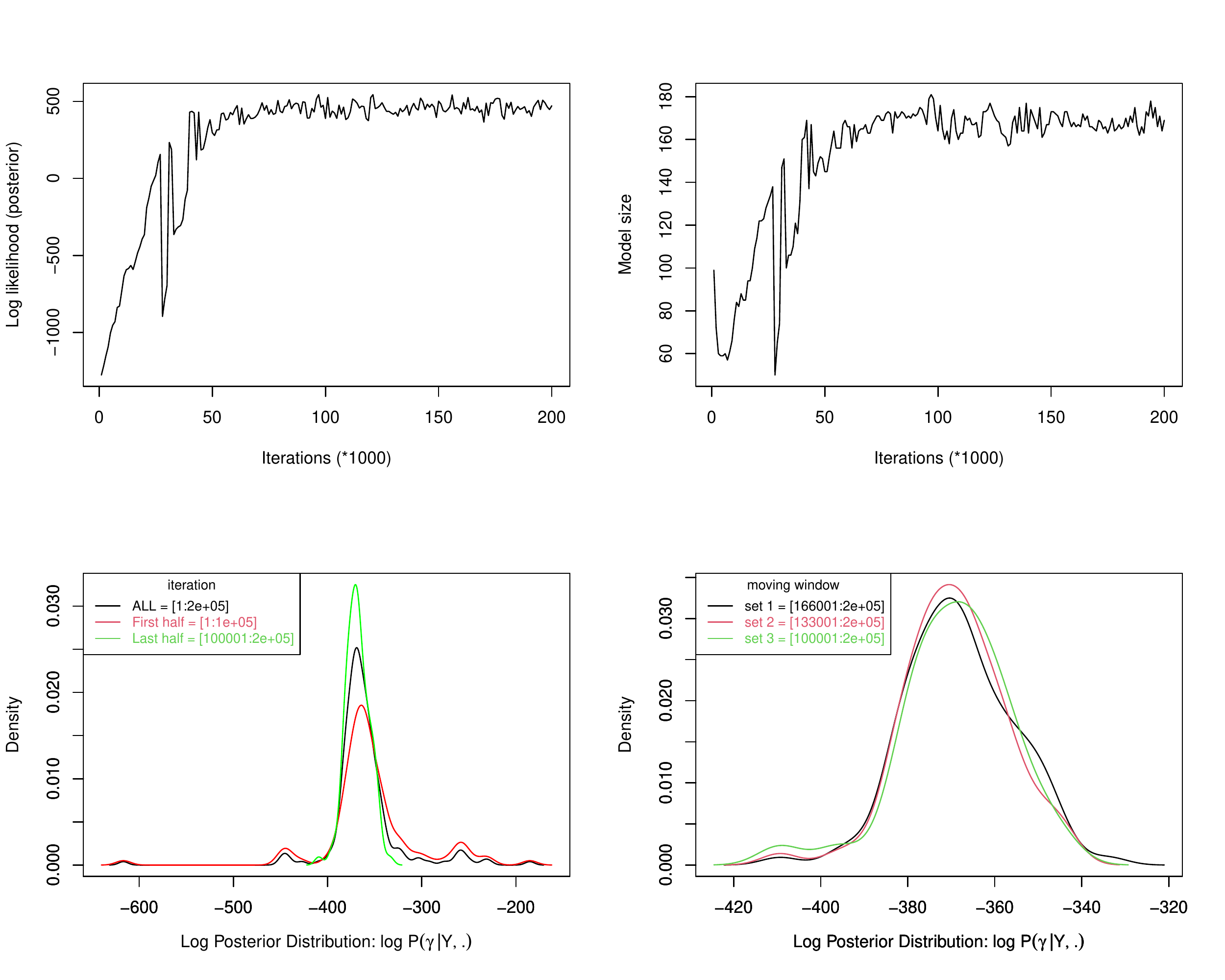}
\caption{Diagnostic plots of the MCMC sampler by \code{plot(fit, estimator = "logP", type = "diagnostics")}.}
\label{MCMCdiag}
\end{figure}

In order to investigate the behaviour of the MCMC sampler, the top two panels of Figure \ref{MCMCdiag} show the trace plots of the loglikelihood and model size, i.e., the total number of selected predictors. We observe that the Markov chain seems to start sampling from the correct distribution after ca. 50,000 iterations. The bottom panels of Figure \ref{MCMCdiag} indicate that the log posterior distribution of the latent indicator variable $\bm\Gamma$  is stable for the last half of the chains after substracting the burn-in length.

\begin{Schunk}
\begin{Sinput}
R> plot(fit, estimator = "logP", type = "diagnostics")
\end{Sinput}
\end{Schunk}

We finish this example analysis by detaching the eQTL example data set.

\begin{Schunk}
\begin{Sinput}
R> detach(exampleEQTL)
\end{Sinput}
\end{Schunk}

\clearpage 
\subsection{The Genomics of Drug Sensitivity in Cancer data}

In this section we analyse a subset of the Genomics of Drug Sensitivity in Cancer (GDSC) dataset from a large-scale pharmacogenomic study \citep{Yang2013, Garnett2012}. We analyse the pharmacological profiling of $n=499$ cell lines from $p_0=13$ different tissue types for $s=7$ cancer drugs. The sensitivity of the cell lines to each of the drugs was summarised by the $\log($IC$_{50}$) values estimated from \textit{in vitro} dose response experiments. The cell lines are characterised by $p_1=343$ selected gene expression features (GEX), $p_2=426$ genes affected by copy number variations (CNV) and $p_3=68$ genes with point mutations (MUT). The data sets were downloaded from \url{ftp://ftp.sanger.ac.uk/pub4/cancerrxgene/releases/release-5.0/} and processed as described in \code{help("exampleGDSC")}. Gene expression features are log-transformed.

\cite{Garnett2012} provide the target genes or pathways for all drugs. The aim of this study was to identify molecular characteristics that help predict the response of a cell line to a particular drug. Because many of the drugs share common targets and mechanisms of action, the response of cell lines to many of the drugs is expected to be correlated. Therefore, a multivariate model seems appropriate:

$$\mathbf{Y}_{\text{drugs}} = \mathbf{X}_{\text{tissues}}\bm{B}_0 + \mathbf{X}_{\text{GEX}}\bm{B}_1 + \mathbf{X}_{\text{CNV}}\bm{B}_2 + \mathbf{X}_{\text{MUT}}\bm{B}_3 + \mathbf{U}_{\text{error}},$$
where the elements of $\bm{B}_0$ and nonzero elements of $\bm{B}_1$, $\bm{B}_2$ and $\bm{B}_3$ are independent and identically distributed with the prior $\mathcal{N}(0, w)$.

\begin{figure}[tph]
\centering
\includegraphics[height=5cm,keepaspectratio]{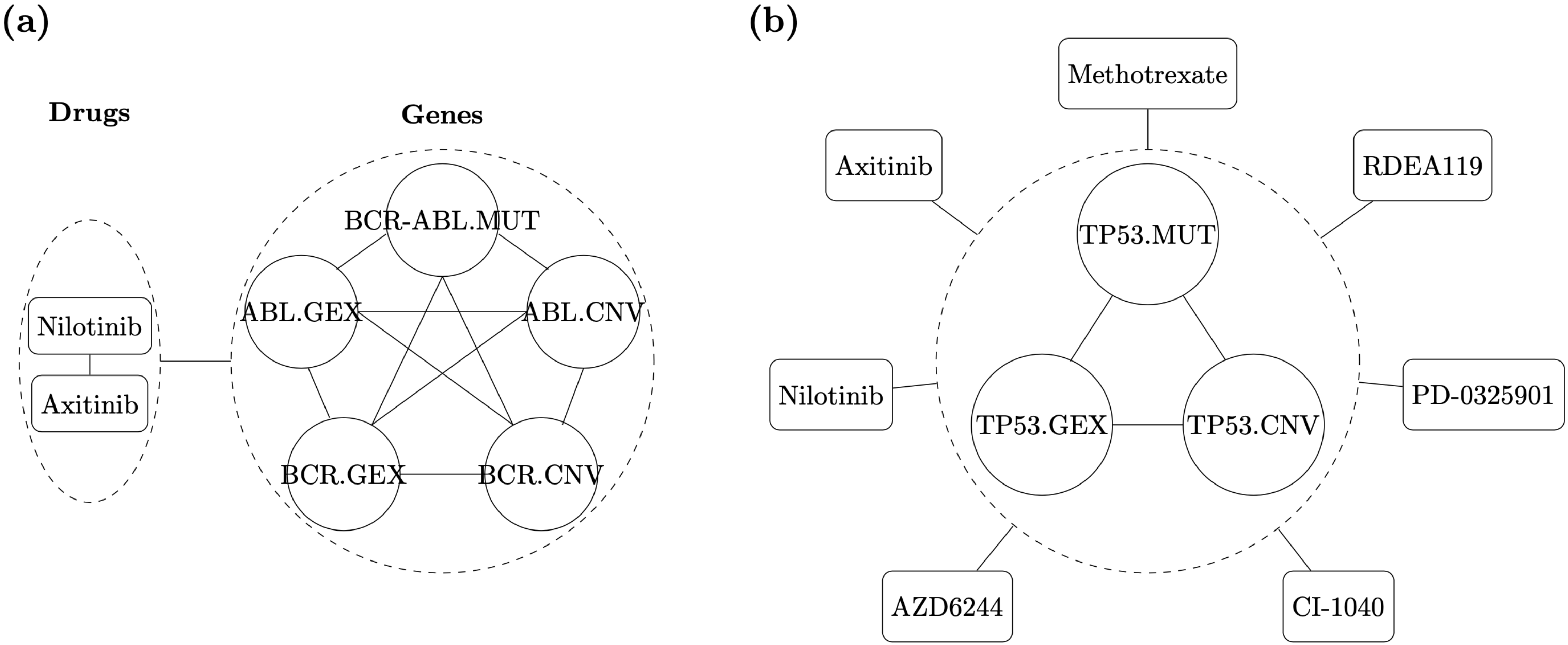}
\caption{Illustration of the relationship between drugs and a group of related genes. The left panel is for the Bcr-Abl fusion gene and the corresponding related genes. The right panel is for all drugs and gene TP35 as one example with features representing all three data sources. The names with suffix ".GEX", ".CNV" and ".MUT" are features of expression, copy number variation and mutation, respectively.}
\label{DrugGenePathway}
\end{figure}

We may know the biological relationships within and between drugs and molecular features, so that the MRF prior (\ref{MRFprior}) can be used to learn the above multivariate model well. In our example, we know that the four drugs RDEA119, PD-0325901, CI-1040 and AZD6244 are MEK inhibitors which affect the MAPK/ERK pathway. Drugs Nilotinib and Axitinib are Bcr-Abl tyrosine kinase inhibitors which inhibit the mutated BCR-ABL gene. Finally, the drug Methotrexate is a chemotherapy agent and general immune system suppressant, which is not associated with a particular molecular target gene or pathway. For the target genes (and genes in target pathways) we consider all characteristics (GEX, CNV, MUT) available in our data set as being potentially associated. Based on this information, we construct edge potentials for the MRF prior:

\begin{itemize}
	\item edges between all features representing genes in the MAPK/ERK pathway and the four MEK inhibitors;
	\item edges between all features representing the Bcr-Abl fusion gene and the two Bcr-Abl inhibitors, see illustration in Figure \ref{DrugGenePathway}(a);
	\item edges between all features from different data sources (i.e., GEX, CNV and MUT) representing a gene and all drugs , see illustration in Figure \ref{DrugGenePathway}(b).
\end{itemize}

By matching the selected genes with the gene set of the MAPK/ERK pathway from the KEGG database, 57 features are considered to be connected to the four MEK inhibitors. The two genes (i.e., BCR and ABL) representing the Bcr-Abl fusion are connected with five features in the data set, which are BCR-ABL mutation, BCR gene expression, BCR copy number variation, ABL gene expression and ABL copy number variation (Figure \ref{DrugGenePathway}(a)). In addition, there are 347 small feature groups representing the different available data sources for each of the genes in the data set, which are potentially connected to all drugs. Figure \ref{DrugGenePathway} (a) illustrates the edges between drugs Nilotinib, Axitinib and the related genes of the Bcr-Abl fusion gene, and Figure \ref{DrugGenePathway} (b) uses the TP53 gene as an example for how the different data sources representing a gene are related to each drug, thus linking the data sources together. Based on this information, we construct the matrix $G$ for the MRF prior.

First, we load and attach the data. Note that in this example, we illustrate the use of the specific plot functions \code{plotEstimator()}, \code{plotGraph()} and \code{plotNetwork()}, which are called directly here rather than via the generic \code{plot()} function as in the examples above.
\begin{Schunk}
\begin{Sinput}
R> data("exampleGDSC", package = "BayesSUR")
R> attach(exampleGDSC)
\end{Sinput}
\end{Schunk}

The following code chunk will run the MCMC sampler to fit the model. This represents a full analysis, which might take several hours to run with the chosen MCMC parameter values (\code{nIter=200000, nChains=6, burnin=100000}) and no parallelisation (\code{maxThreads=1} by default). Approximate results for an initial assessment of the model can be achieved with much shorter MCMC runs. Note that we use the \code{X\_0} argument for the thirteen cancer tissue types, which are included in the model as mandatory predictors that are always selected.
\begin{Schunk}
\begin{Sinput}
R> hyperpar <- list(mrf_d = -3, mrf_e = 0.2)
R> set.seed(6437)
R> tic("Time of model fitting")
R> fit <- BayesSUR(data = data, Y = blockList[[1]], X_0 = blockList[[2]], 
+                  X = blockList[[3]], outFilePath = "results/", 
+                  nIter = 200000, burnin = 100000, nChains = 6, 
+                  covariancePrior = "HIW", gammaPrior = "MRF", 
+                  hyperpar = hyperpar, mrfG = mrfG)
\end{Sinput}
\end{Schunk}
\vspace{-8.8mm}
\begin{Schunk}
\begin{Sinput}
R> toc()
\end{Sinput}
\end{Schunk}

\code{Time of model fitting: 7468.874 sec elapsed}\\

\begin{figure}[tph]
\begin{center}
  \includegraphics[height=6cm,keepaspectratio]{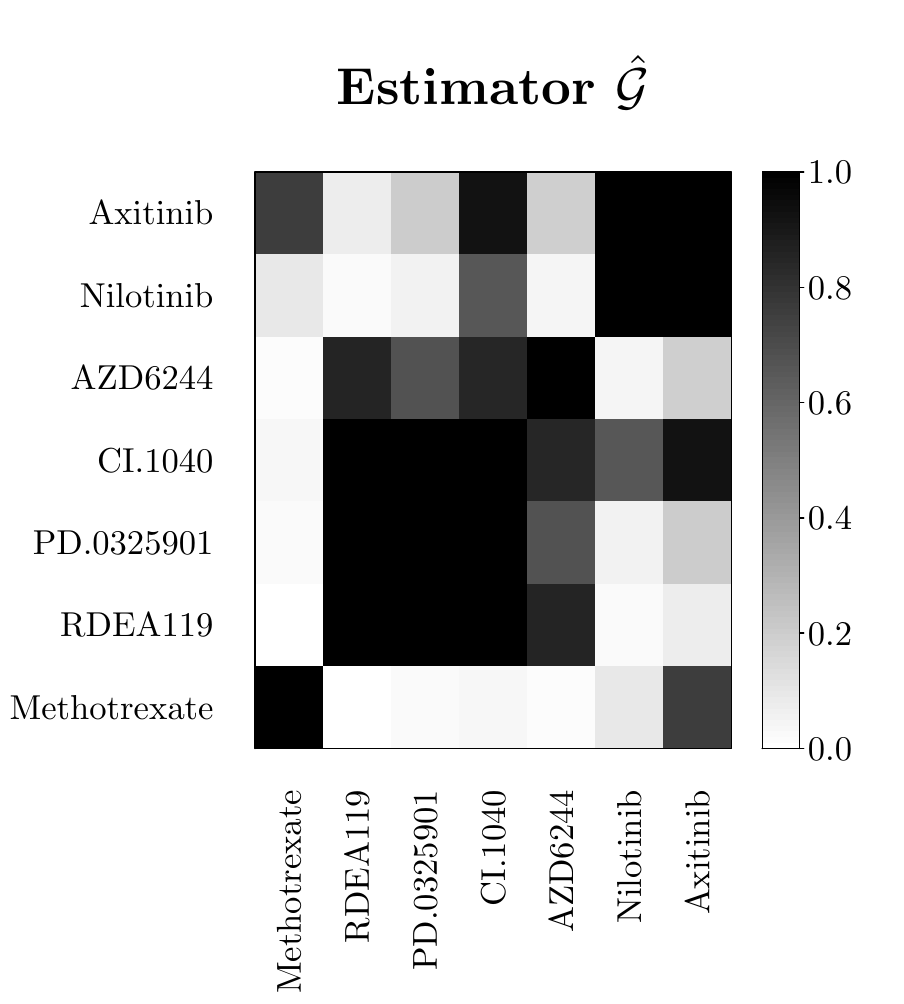}
\qquad
  \includegraphics[height=7cm,keepaspectratio]{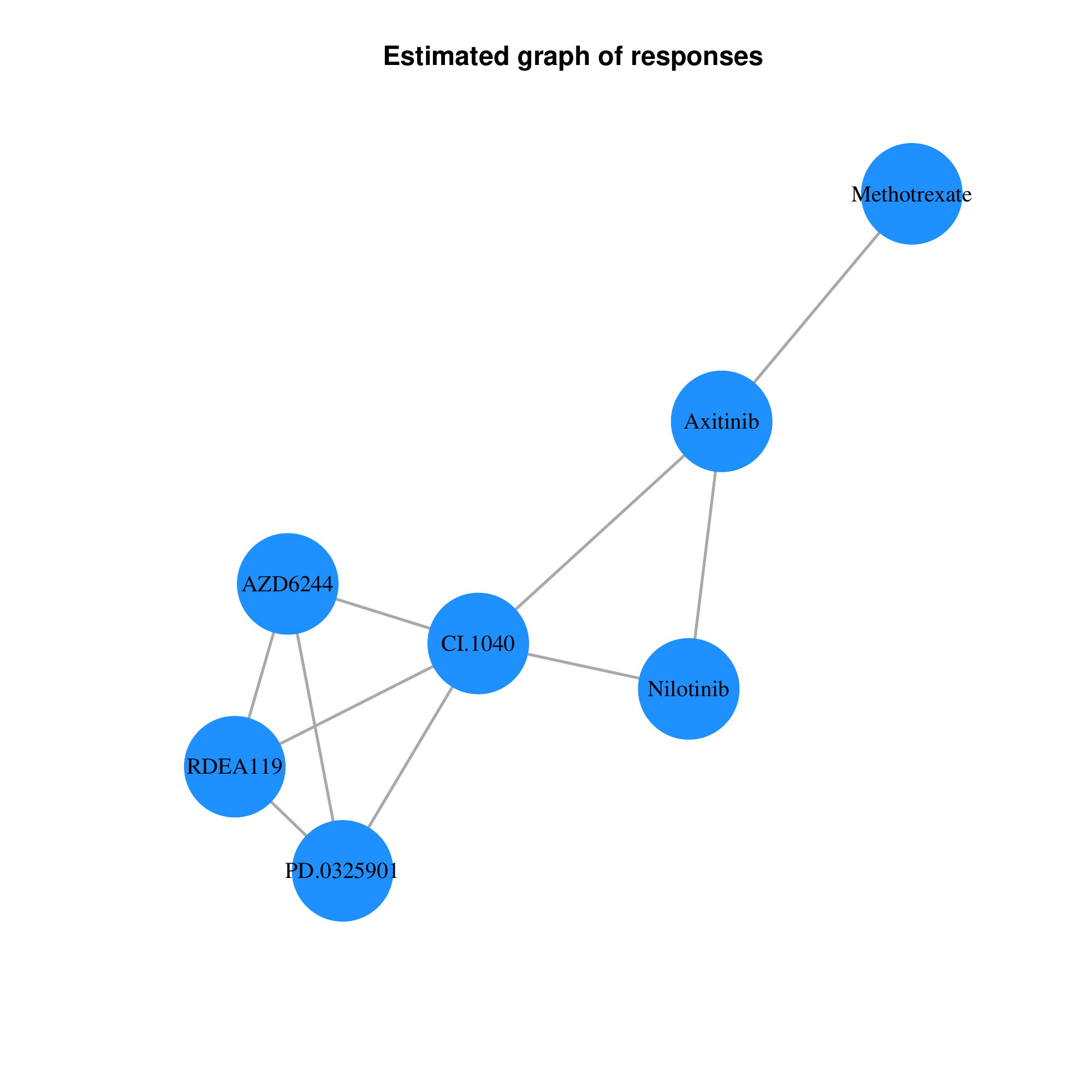}
\caption{Estimated structure of the seven drugs $\hat{\mathcal{G}}$. Their associations as visualised in the right panel are based on $\hat{\mathcal{G}}$ thresholded at 0.5. Figures created with \code{plotEstimator()} (left) and \code{plotGraph()} (right).
\label{ResponseGraphGDSC}}
\end{center}
\end{figure}

After fitting an SSUR model with the MRF prior, the structure of the seven drugs, $\mathcal{G}$, has been learned as illustrated in Figure \ref{ResponseGraphGDSC}, where edges between two drugs $k$ and $k'$ indicate that $\hat{\mathcal{G}}_{kk'} > 0.5$. All expected associations between the drugs within each drug group are found, but some additional connections are also identified: there are edges between Axitinib and Methotrexate and between CI-1040 and both Nilotinib and Axitinib.

\begin{Schunk}
\begin{Sinput}
R> plotEstimator(fit, estimator = "Gy", name.responses = c("Methotrexate", 
+                "RDEA119", "PD.0325901",  "CI.1040", "AZD6244", "Nilotinib",
+                "Axitinib"), fig.tex = TRUE, output = "ResponseGraphGDSC1")
\end{Sinput}
\end{Schunk}
\begin{Schunk}
\begin{Sinput}
R> plotGraph(fit, estimator = "Gy")
\end{Sinput}
\end{Schunk}

\begin{Schunk}
\begin{Sinput}
R> plotNetwork(fit, estimator = c("gamma","Gy"), label.predictor = "", 
+              name.predictors = "Genes", name.responses = "Drugs", 
+              nodesizePredictor = 2)
\end{Sinput}
\end{Schunk}
\begin{figure}[tph]
\centering
\includegraphics[height=10cm,keepaspectratio]{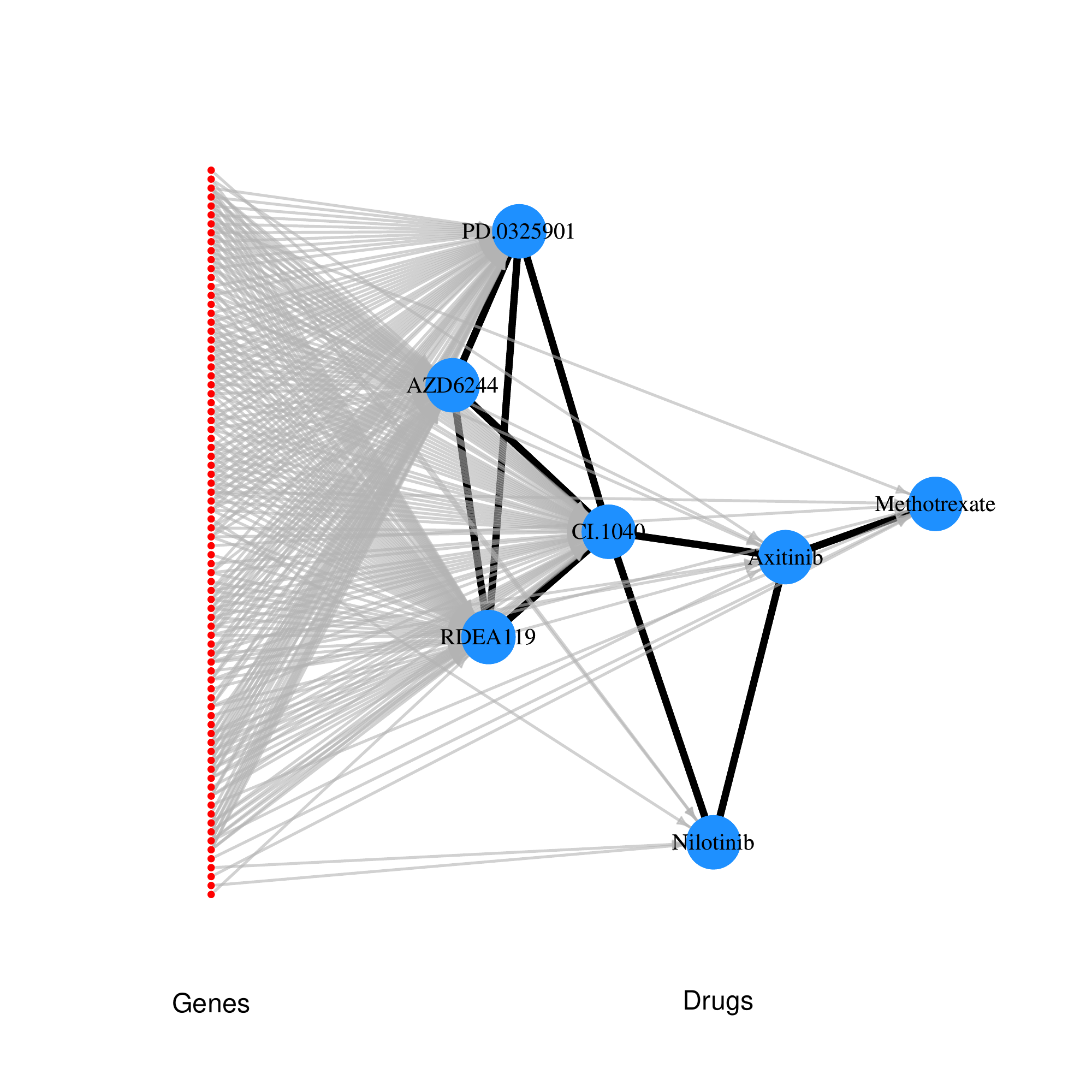}
\caption{Estimated network between the seven drugs and selected genes based on thresholds 0.5. Figure created with \code{plotNetwork()}.}
\label{NetworkFull}
\end{figure}

The estimated relationships between the drugs and genes are displayed in Figure \ref{NetworkFull}. There are 259 of all 5859 coefficients selected in total when thresholding $\hat{\bm\Gamma}$ at 0.5. This results in 82 molecular features being selected for at least one of the drugs, 7 for Methotrexate, 69 for the four MEK inhibitors and 11 for the two Bcr-Abl tyrosine kinase inhibitors.

\begin{figure}[tph]
\centering
\includegraphics[height=8cm,keepaspectratio]{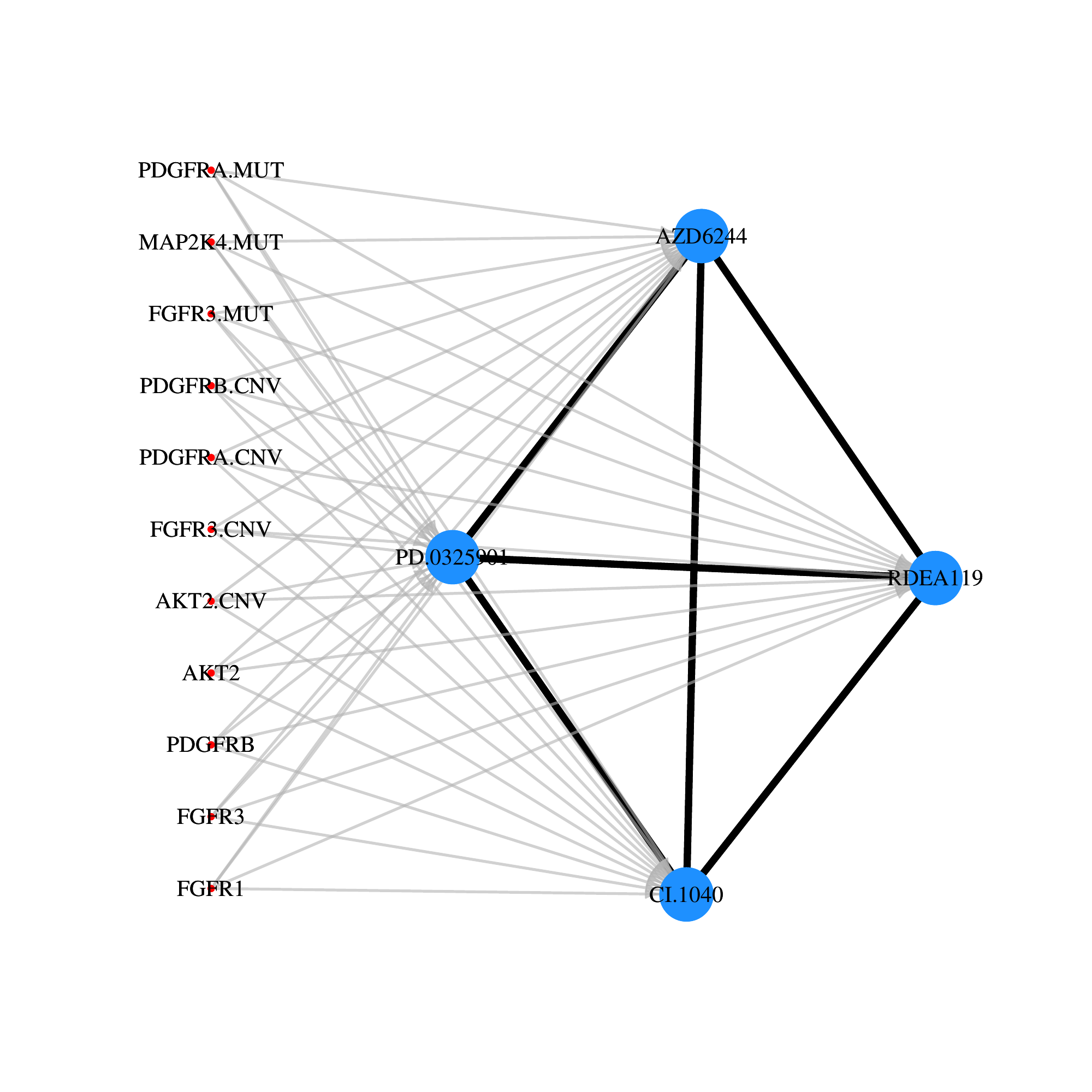}
\caption{Estimated network between the MEK inhibitors and selected target genes based on thresholds 0.5. Figure created with \code{plotNetwork()}.}
\label{Network1}
\end{figure}

\begin{figure}[tph]
\centering
\includegraphics[height=5cm,keepaspectratio]{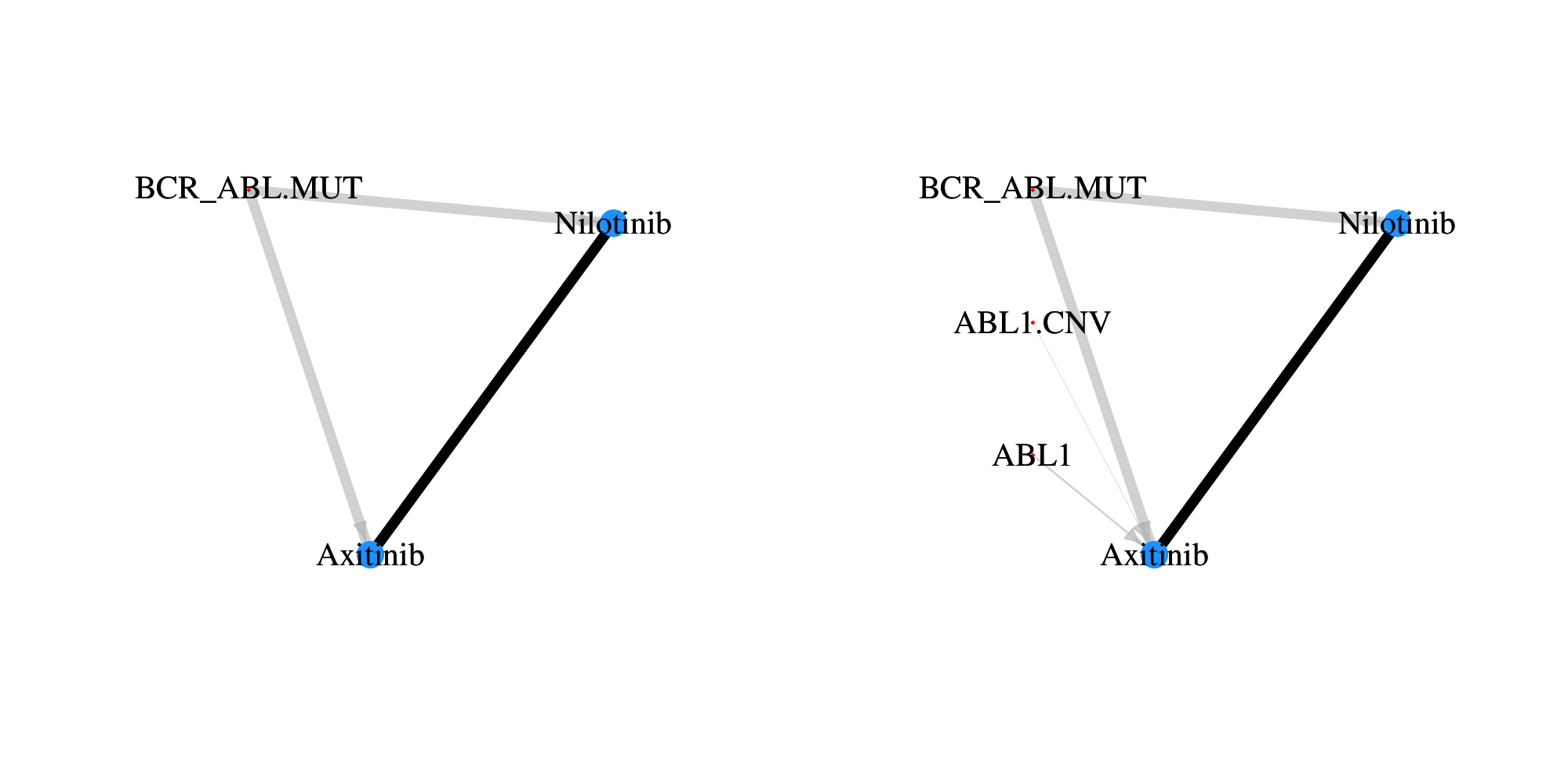}
\caption{Estimated network between the Bcr-Abl inhibitors and selected target genes. The left panel is based on threshold on $\hat{\bm\Gamma}$ of 0.5 while the right panel is based on threshold 0.01. Both panels use a threshold on $\hat{\mathcal{G}}$ of 0.5. The edges are weighted by the corresponding inclusion probabilities, if they are greater than the specified thresholds. Figures created with \code{plotNetwork()}.}
\label{Network2}
\end{figure}

Network substructures of interest can also be selected and visualised individually, since the user can specify, which response variables (drugs) and which input variables (molecular features) to include in a figure. For example, Figures \ref{Network1} and \ref{Network2} show the estimated network representations of the two groups of drugs, respectively.
\begin{Schunk}
\begin{Sinput}
R> data("targetGene", package = "BayesSUR")
R> plotNetwork(fit, estimator = c("gamma","Gy"), 
+        includeResponse = c("RDEA119", "PD.0325901", "CI.1040", "AZD6244"),
+        includePredictor = names(targetGene$group1))
\end{Sinput}
\end{Schunk}

In addition, Figure \ref{Network2} illustrates, how one can customise how to display the edges between input and response variables to visualise the strength of the association between nodes. In particular, one can either simply use a threshold, e.g.,  0.5, to show all edges with marginal posterior inclusion probabilities larger than the threshold equally (left panel), or the width of edges (greater than the specified threshold) can be weighted by the corresponding inclusion probability (right panel).

\begin{Schunk}
\begin{Sinput}
R> layout(matrix(1:2, ncol = 2))
R> plotNetwork(fit, estimator = c("gamma","Gy"), edge.weight = TRUE,
+              includeResponse = c("Nilotinib", "Axitinib"), 
+              includePredictor = names(targetGene$group2)) 
R> plotNetwork(fit, estimator = c("gamma","Gy"), 
+              edge.weight = TRUE, PmaxPredictor = 0.01, 
+              includeResponse = c("Nilotinib", "Axitinib"), 
+              includePredictor =names(targetGene$group2)) 
\end{Sinput}
\end{Schunk}

\section{Conclusion} \label{sec:conclusion}

The \pkg{BayesSUR} package presents a series of multivariate Bayesian variable selection models, for which the ESS algorithm is employed for posterior inference over the model space. It provides a unified \proglang{R} package and a consistent interface for the \proglang{C++} implementations of individual models. The package supports all combinations of the covariance priors and variable selection priors from Section \ref{sec:models} in the Bayesian HRR and SUR model frameworks. This includes the MRF prior on the latent indicator variables to allow the user to make use of prior knowledge of the relationships between both response variables and predictors. To overcome the computational cost for datasets with large numbers of input variables, parallel processing is also implemented with respect to multiple chains, and for calculation of likelihoods of parameters and samples, although the MCMC algorithm itself is still challenging to be parallelised. We demonstrated the modelling aspects of variable selection and structure recovery to identify relationships between multivariate (potentially high-dimensional) responses as well as between responses and high-dimensional predictors, by applying the package to a simulated eQTL dataset and to pharmacogenomic data from the GDSC project. 

Possible extensions of the \proglang{R} package include the implementation of different priors to introduce even more flexibility in the modelling choices. In particular, the $g$-prior could be considered for the regression coefficients matrix $\bm{B}$ \citep{Bottolo2010, Richardson2011, Lewin2016}, whereas currently only the independence prior is available. In addition, the spike-and-slab prior on the covariance matrix $C$ \citep{Wang2015, Banerjee2015, Deshpande2019} might be useful, or the horseshoe prior on the latent indicator variable $\bm{\Gamma}$, which was recently implemented in the multivariate regression setup by \cite{Ruffieux2020}.

\section{Acknowledgements} \label{sec:acknowledgement}
$^*$A. Lewin and M. Zucknick are joint last authors. The authors thank the editors and the two referees for helpful suggestions. The authors declare no conflicts of interest. This work was made possible through funding from the Faculty of Medicine, University of Oslo (ZZ, MZ), Research Council of Norway project No. 237718 ``Big Insight'' (ZZ), European Union Horizon 2020 grant agreements No. 847912 ``RESCUER'' (MZ, SR) and No. 633595 ``DynaHealth'' (AL), UK Medical Research Council grants MR/M013138/1 (MB, AL, LB, SR) and MC\_UU\_00002/10 (SR), NIHR Cambridge BRC (SR), BHF-Turing Cardiovascular Data Science Awards 2017 (LB) and The Alan Turing Institute under UK Engineering and Physical Sciences Research Council grant EP/N510129/1 (LB).


\bibliography{jss3833}

\begin{thebibliography}{37}
\newcommand{\enquote}[1]{``#1''}
\providecommand{\natexlab}[1]{#1}
\providecommand{\url}[1]{\texttt{#1}}
\providecommand{\urlprefix}{URL }
\expandafter\ifx\csname urlstyle\endcsname\relax
  \providecommand{\doi}[1]{doi:\discretionary{}{}{}#1}\else
  \providecommand{\doi}{doi:\discretionary{}{}{}\begingroup
  \urlstyle{rm}\Url}\fi
\providecommand{\eprint}[2][]{\url{#2}}

\bibitem[{Banerjee(2008)}]{Banerjee2008}
Banerjee S (2008).
\newblock \enquote{Bayesian linear model: Gory details.}
\newblock
  \urlprefix\url{http://www.biostat.umn.edu/~ph7440/pubh7440/BayesianLinearModelGoryDetails.pdf}.

\bibitem[{Banerjee and Ghosal(2015)}]{Banerjee2015}
Banerjee S, Ghosal S (2015).
\newblock \enquote{Bayesian structure learning in graphical models.}
\newblock \emph{Journal of Multivariate Analysis}, \textbf{136}, 147--162.
\newblock \doi{10.1016/j.jmva.2015.01.015}.

\bibitem[{Banterle \emph{et~al.}(2018)Banterle, Bottolo, Richardson,
  Ala-Korpela, Jarvelin, and Lewin}]{Banterle2018}
Banterle M, Bottolo L, Richardson S, Ala-Korpela M, Jarvelin MR, Lewin A
  (2018).
\newblock \enquote{Sparse variable and covariance selection for
  high-dimensional seemingly unrelated Bayesian regression.}
\newblock \emph{bioRxiv}, p. 467019.
\newblock \doi{10.1101/467019}.

\bibitem[{Barbieri and Berger(2004)}]{Barbieri2004}
Barbieri MM, Berger JO (2004).
\newblock \enquote{Optimal predictive model selection.}
\newblock \emph{The Annals of Statistics}, \textbf{32}(3), 870--897.

\bibitem[{Barretina \emph{et~al.}(2012)Barretina, Caponigro, Stransky,
  Venkatesan, Margolin, Kim, Wilson, Lehar, Kryukov, Sonkin, and
  \emph{others}}]{Barretina2012}
Barretina J, Caponigro G, Stransky N, Venkatesan K, Margolin A, Kim S, Wilson
  C, Lehar J, Kryukov G, Sonkin D, \emph{others} (2012).
\newblock \enquote{The Cancer Cell Line Encyclopedia enables predictive
  modelling of anticancer drug sensitivity.}
\newblock \emph{Nature}, \textbf{483}(7391), 603--607.
\newblock \doi{10.1038/nature11003}.

\bibitem[{Bhadra and Mallick(2013)}]{Bhadra2013}
Bhadra A, Mallick BK (2013).
\newblock \enquote{Joint High-Dimensional Bayesian Variable and Covariance
  Selection with an Application to eQTL Analysis.}
\newblock \emph{Biometrics}, \textbf{69}(2), 447--457.
\newblock \doi{10.1111/biom.12021}.

\bibitem[{Bottolo \emph{et~al.}(2011)Bottolo, Petretto, Blankenberg, Cambien,
  Cook, Tiret, and Richardson}]{Bottolo2011}
Bottolo L, Petretto E, Blankenberg S, Cambien F, Cook SA, Tiret L, Richardson S
  (2011).
\newblock \enquote{Bayesian detection of expression quantitative trait loci
  hot-spots.}
\newblock \emph{Genetics}, \textbf{189}(4), 1449--1459.
\newblock \doi{10.1534/genetics.111.131425}.

\bibitem[{Bottolo and Richardson(2010)}]{Bottolo2010}
Bottolo L, Richardson S (2010).
\newblock \enquote{Evolutionary Stochastic Search for Bayesian Model
  Exploration.}
\newblock \emph{Bayesian Analysis}, \textbf{5}(3), 583--618.
\newblock \doi{10.1214/10-BA523}.

\bibitem[{Carvalho \emph{et~al.}(2007)Carvalho, Massam, and
  West}]{Carvalho2007}
Carvalho CM, Massam H, West M (2007).
\newblock \enquote{Simulation of Hyper-Inverse Wishart Distributions in
  Graphical Models.}
\newblock \emph{Biometrika}, \textbf{94}(3), 647--659.
\newblock \doi{10.1093/biomet/asm056}.

\bibitem[{Cs{\'a}rdi and Nepusz(2006)}]{Csardi_2006}
Cs{\'a}rdi G, Nepusz T (2006).
\newblock \enquote{The igraph software package for complex network research.}
\newblock \emph{InterJournal, Complex Systems}, \textbf{1695}(5), 1--9.
\newblock \urlprefix\url{http://igraph.sf.net}.

\bibitem[{Deshpande \emph{et~al.}(2019)Deshpande, Ro{\v{c}}kov{\'a}, and
  George}]{Deshpande2019}
Deshpande SK, Ro{\v{c}}kov{\'a} V, George EI (2019).
\newblock \enquote{Simultaneous variable and covariance selection with the
  multivariate spike-and-slab lasso.}
\newblock \emph{Journal of Computational and Graphical Statistics},
  \textbf{28}(4), 921--931.
\newblock \doi{10.1080/10618600.2019.1593179}.

\bibitem[{Eddelbuettel and Fran\c{c}ois(2011)}]{Edelbuettel_2011}
Eddelbuettel D, Fran\c{c}ois R (2011).
\newblock \enquote{{Rcpp}: Seamless {R} and {C++} Integration.}
\newblock \emph{Journal of Statistical Software}, \textbf{40}(8), 1--18.
\newblock \doi{10.18637/jss.v040.i08}.

\bibitem[{Eddelbuettel and Sanderson(2014)}]{Edelbuettel_2014}
Eddelbuettel D, Sanderson C (2014).
\newblock \enquote{RcppArmadillo: Accelerating R with high-performance C++
  linear algebra.}
\newblock \emph{Computational Statistics and Data Analysis}, \textbf{71},
  1054--1063.
\newblock \doi{10.1016/j.csda.2013.02.005}.

\bibitem[{Garnett \emph{et~al.}(2012)Garnett, Edelman, Heidorn, Greenman,
  Dastur, Lau, Greninger, Thompson, Luo, Soares, Liu, and
  \emph{others}}]{Garnett2012}
Garnett M, Edelman E, Heidorn S, Greenman C, Dastur A, Lau K, Greninger P,
  Thompson I, Luo X, Soares J, Liu Q, \emph{others} (2012).
\newblock \enquote{Systematic identification of genomic markers of drug
  sensitivity in cancer cells.}
\newblock \emph{Nature}, \textbf{483}(7391), 570--575.
\newblock \doi{10.1038/nature11005}.

\bibitem[{Gelfand(1996)}]{Gelfand1996}
Gelfand A (1996).
\newblock \enquote{Model Determination Using Sampling Based Method.}
\newblock In W~Gilks, S~Richardson, D~Spiegelhalter (eds.), \emph{Markov Chain
  Monte Carlo in Practice}, chapter~9, pp. 145--161. Chapman \& Hall, Boca
  Raton, FL.

\bibitem[{Gray and Mills(2015)}]{Gray2015}
Gray JW, Mills GB (2015).
\newblock \enquote{Large-Scale Drug Screens Support Precision Medicine.}
\newblock \emph{Cancer Discovery}, \textbf{5}(11), 1130--1132.
\newblock \doi{10.1158/2159-8290.CD-15-0945}.

\bibitem[{Green and Thomas(2013)}]{Green2013}
Green PJ, Thomas A (2013).
\newblock \enquote{Sampling decomposable graphs using a Markov chain on
  junction trees.}
\newblock \emph{Biometrika}, \textbf{100}(1), 91--110.
\newblock \doi{10.1093/biomet/ass052}.

\bibitem[{Holmes \emph{et~al.}(2002)Holmes, Denison, and Mallick}]{Holmes2002}
Holmes C, Denison D, Mallick B (2002).
\newblock \enquote{Accounting for Model Uncertainty in Seemingly Unrelated
  Regressionsa.}
\newblock \emph{Journal of Computational and Graphical Statistics},
  \textbf{11}(3), 533--551.
\newblock \doi{10.1198/106186002475}.

\bibitem[{Jia and Xu(2007)}]{Jia2007}
Jia Z, Xu S (2007).
\newblock \enquote{Mapping Quantitative Trait Loci for Expression Abundance.}
\newblock \emph{Genetics}, \textbf{176}(1), 611--623.
\newblock \doi{10.1534/genetics.106.065599}.

\bibitem[{Jones \emph{et~al.}(2005)Jones, Carvalho, Dobra, Hans, Carter, and
  West}]{Jones2005}
Jones B, Carvalho C, Dobra A, Hans C, Carter C, West M (2005).
\newblock \enquote{Experiments in Stochastic Computation for High-Dimensional
  Graphical Models.}
\newblock \emph{Statistical Science}, \textbf{20}(4), 388--400.
\newblock \doi{10.1214/088342305000000304}.

\bibitem[{Lee \emph{et~al.}(2017)Lee, Tadesse, Baccarelli, Schwartz, and
  Coull}]{Lee2017}
Lee KH, Tadesse MG, Baccarelli AA, Schwartz J, Coull BA (2017).
\newblock \enquote{Multivariate Bayesian variable selection exploiting
  dependence structure among outcomes: Application to air pollution effects on
  DNA methylation.}
\newblock \emph{Biometrics}, \textbf{73}(1), 232--241.
\newblock \doi{10.1111/biom.12557}.

\bibitem[{Lewin \emph{et~al.}(2015)Lewin, Saadi, Peters, Moreno-Moral, Lee,
  Smith, Petretto, Bottolo, and Richardson}]{Lewin2016}
Lewin A, Saadi H, Peters JE, Moreno-Moral A, Lee JC, Smith KG, Petretto E,
  Bottolo L, Richardson S (2015).
\newblock \enquote{MT-HESS: an efficient Bayesian approach for simultaneous
  association detection in OMICS datasets, with application to eQTL mapping in
  multiple tissues.}
\newblock \emph{Bioinformatics}, \textbf{32}(4), 523--532.
\newblock \doi{10.1093/bioinformatics/btv568}.

\bibitem[{Liang and Wong(2000)}]{Liang2000}
Liang F, Wong WH (2000).
\newblock \enquote{Evolutionary Monte Carlo: Applications to $C_{p}$ model
  sampling and change point problem.}
\newblock \emph{Statistica Sinica}, \textbf{10}(2), 317--342.
\newblock \urlprefix\url{http://www.jstor.org/stable/24306722}.

\bibitem[{Liquet \emph{et~al.}(2016)Liquet, Bottolo, Campanella, Richardson,
  and Chadeau-Hyam}]{Liquet2016}
Liquet B, Bottolo L, Campanella G, Richardson S, Chadeau-Hyam M (2016).
\newblock \enquote{R2GUESS: A Graphics Processing Unit-Based R Package for
  Bayesian Variable Selection Regression of Multivariate Responses.}
\newblock \emph{Journal of Statistical Software, Articles}, \textbf{69}(2),
  1--32.
\newblock ISSN 1548-7660.
\newblock \doi{10.18637/jss.v069.i02}.

\bibitem[{Liquet \emph{et~al.}(2017)Liquet, Mengersen, Pettitt, and
  Sutton}]{Liquet2017}
Liquet B, Mengersen K, Pettitt AN, Sutton M (2017).
\newblock \enquote{Bayesian Variable Selection Regression of Multivariate
  Responses for Group Data.}
\newblock \emph{Bayesian Analysis}, \textbf{12}(4), 1039--1067.
\newblock \doi{10.1214/17-BA1081}.

\bibitem[{Mohammadi and Wit(2019)}]{Mohammadi2019}
Mohammadi R, Wit E (2019).
\newblock \enquote{BDgraph: An R package for Bayesian structure learning in
  graphical models.}
\newblock \emph{Journal of Statistical Software}, \textbf{89}(3), 1--30.
\newblock ISSN 1548-7660.
\newblock \doi{10.18637/jss.v089.i03}.

\bibitem[{Petretto \emph{et~al.}(2010)Petretto, Bottolo, Langley, Heinig,
  Mcdermott-Roe, Sarwar, Pravenec, Hubner, Aitman, Cook, and
  Richardson}]{Petretto2010}
Petretto E, Bottolo L, Langley SR, Heinig M, Mcdermott-Roe C, Sarwar R,
  Pravenec M, Hubner N, Aitman TJ, Cook SA, Richardson S (2010).
\newblock \enquote{New Insights into the Genetic Control of Gene Expression
  using a Bayesian Multi-tissue Approach.}
\newblock \emph{PLoS Computational Biology}, \textbf{6}(4), e1000737.
\newblock \doi{10.1371/journal.pcbi.1000737}.

\bibitem[{Richardson \emph{et~al.}(2011)Richardson, Bottolo, and
  Rosenthal}]{Richardson2011}
Richardson S, Bottolo L, Rosenthal JS (2011).
\newblock \enquote{Bayesian Models for Sparse Regression Analysis of High
  Dimensional Data.}
\newblock \emph{Bayesian Statistics 9}, pp. 539--568.
\newblock \doi{10.1093/acprof:oso/9780199694587.001.0001}.

\bibitem[{Ruffieux \emph{et~al.}(2020)Ruffieux, Davison, Hager, Inshaw,
  Fairfax, Richardson, and Bottolo}]{Ruffieux2020}
Ruffieux H, Davison AC, Hager J, Inshaw J, Fairfax BP, Richardson S, Bottolo L
  (2020).
\newblock \enquote{A global-local approach for detecting hotspots in
  multiple-response regression.}
\newblock \emph{The Annals of Applied Statistics}, \textbf{14}(2), 905--928.
\newblock \doi{10.1214/20-AOAS1332}.

\bibitem[{Schwender and Fritsch(2012)}]{Schwender2018}
Schwender H, Fritsch A (2012).
\newblock \enquote{scrime: Analysis of high-dimensional categorical data such
  as SNP data.}
\newblock \emph{R package version 1.3.5}.
\newblock \urlprefix\url{https://CRAN.R-project.org/package=scrime}.

\bibitem[{Stingo \emph{et~al.}(2011)Stingo, Chen, Tadesse, and
  Vannucci}]{Stingo2011}
Stingo FC, Chen YA, Tadesse MG, Vannucci M (2011).
\newblock \enquote{Incorporating biological information into linear models: A
  Bayesian approach to the selection of pathways and genes.}
\newblock \emph{The Annals of Applied Statistics}, \textbf{5}(3), 1978--2002.

\bibitem[{Uhler \emph{et~al.}(2018)Uhler, Lenkoski, and Richards}]{Uhler2018}
Uhler C, Lenkoski A, Richards D (2018).
\newblock \enquote{Exact Formulas for the Normalizing Constants of Wishart
  Distributions for Graphical Models.}
\newblock \emph{The Annals of Statistics}, \textbf{46}(1), 90--118.
\newblock \doi{10.1214/17-AOS1543}.

\bibitem[{Vehtari \emph{et~al.}(2017)Vehtari, Gelman, and Gabry}]{Vehtari2017}
Vehtari A, Gelman A, Gabry J (2017).
\newblock \enquote{Practical Bayesian model evaluation using leave-one-out
  cross-validation and WAIC.}
\newblock \emph{Statistics and Computing}, \textbf{27}(5), 1413--1432.
\newblock \doi{10.1007/s11222-016-9696-4}.

\bibitem[{Wang(2010)}]{Wang2010}
Wang H (2010).
\newblock \enquote{Sparse Seemingly Unrelated Regression Modelling:
  Applications in Finance and Econometrics.}
\newblock \emph{Comput. Stat. Data Anal.}, \textbf{54}(11), 2866--2877.
\newblock \doi{10.1016/j.csda.2010.03.028}.

\bibitem[{Wang(2015)}]{Wang2015}
Wang H (2015).
\newblock \enquote{Scaling it up: Stochastic search structure learning in
  graphical models.}
\newblock \emph{Bayesian Analysis}, \textbf{10}(2), 351--377.
\newblock \doi{10.1214/14-BA916}.

\bibitem[{Yang \emph{et~al.}(2013)Yang, Soares, Greninger, Edelman, Lightfoot,
  Forbes, Bindal, Beare, Smith, Thompson, Ramaswamy, Futreal, Haber, Stratton,
  Benes, McDermott, and Garnett}]{Yang2013}
Yang W, Soares J, Greninger P, Edelman E, Lightfoot H, Forbes S, Bindal N,
  Beare D, Smith J, Thompson I, Ramaswamy S, Futreal P, Haber D, Stratton M,
  Benes C, McDermott U, Garnett M (2013).
\newblock \enquote{Genomics of Drug Sensitivity in Cancer (GDSC): a resource
  for therapeutic biomarker discovery in cancer cells.}
\newblock \emph{Nucleic Acids Reserch}, \textbf{41}(Database issue), D955--61.
\newblock \doi{10.1093/nar/gks1111}.

\bibitem[{Zhao \emph{et~al.}(2021)Zhao, Banterle, Lewin, and
  Zucknick}]{Zhao2021}
Zhao Z, Banterle M, Lewin A, Zucknick M (2021).
\newblock \enquote{Structured Bayesian variable selection for multiple related
  response variables and high-dimensional predictors.}
\newblock \emph{arXiv preprint arXiv:2101.05899}.
\newblock \urlprefix\url{https://arxiv.org/abs/2101.05899}.

\end{thebibliography}


\clearpage
\section*{Appendix for the elpd} \label{sec:appendix}

Without loss of generality, here we only consider each response variable $\bm{y}$ of the whole response matrix $\mathbf{Y}$. Then the basic linear model is
\begin{align*} \tag{A.1}\label{append.model}
\begin{split}
\bm{y} | X, \beta, \sigma^2 &\sim \mathcal{N}(X\beta, \sigma^2\mathbbm{I}_m), \\
\beta | \sigma^2 &\sim \mathcal{N}(\mu_{\beta}, \sigma^2V_{\beta}), \\
\sigma^2 &\sim \mathcal{IG}(a, b).
\end{split}
\end{align*}
In \cite{Bottolo2011} and the HRR model of this article, $\mu_{\beta}=\bm{0}$ and $V_{\beta}=\mathbbm{I}_p$ for nonzero coefficients.

\subsection*{Appendix 1: Posterior predictive for the HRR model}

From (\ref{append.model}), the joint distribution of $(\beta,\sigma^2)$ is Normal-Inverse-Gamma, i.e.,
$$f(\beta, \sigma^2) = f(\beta |\sigma^2)f(\sigma^2) = \mathcal{N}(\mu_{\beta}, V_{\beta}) \cdot  \mathcal{IG}(a, b) = \mathcal{NIG}(\mu_{\beta}, \sigma^2V_{\beta}, a,b). $$
Further we can know the posterior distribution of $(\beta,\sigma^2)$ is still Normal-Inverse-Gamma $\mathcal{NIG}(\mu_{\beta}^*, V_{\beta}^*, a^*,b^*)$ \citep{Banerjee2008}, where
\begin{align*}
\mu^* &= (V_{\beta}^{-1} + X^\top X)^{-1} (V_{\beta}^{-1}\mu_{\beta} + X^\top y), \\
V^* &= (V_{\beta}^{-1} + X^\top X)^{-1}, \\
a^* &= a + \frac{n}{2}, \\
b^* &= b + \frac{1}{2}( \mu_{\beta}^\top V_{\beta}^{-1} \mu_{\beta} + \bm{y}^\top \bm{y} - \mu^{*\top}V^{*-1}\mu^* ).
\end{align*}

Now we derive the posterior predictive w.r.t. individual response $y_i$.
\begin{align*}
f(y_i |\bm{y}) &= \int f(y_i |\beta, \sigma^2) f(\beta, \sigma^2 | y) d\beta d\sigma^2 \\
&= \int \mathcal{N}(X_i\beta, \sigma^2) \cdot \mathcal{NIG} (\mu^*, V^*, a^*, b^*) d\beta d\sigma^2 \\
&= \int \frac{b^{*a^*}}{ (2\pi)^{\frac{p+1}{2}} \Gamma(a^*) |V^*|^{1/2}} \left(\frac{1}{\sigma^2}\right)^{a^*+\frac{p+1}{2}+1} \times \\ &\ \ \ \ \ \ \exp\left\{-\frac{1}{\sigma^2} \left[ b^*+\frac{1}{2} \left\{ (\beta-\mu^*)^\top V^{*-1}(\beta-\mu^*) + (y_i-X_i\beta)^2 \right\} \right] \right\} d\beta d\sigma^2 \\
&= \int \frac{b^{*a^*}}{ (2\pi)^{\frac{p+1}{2}} \Gamma(a^*) |V^*|^{1/2}} \left(\frac{1}{\sigma^2}\right)^{a^*+\frac{p+1}{2}+1} \times \\ &\ \ \ \ \ \ \exp\left\{-\frac{1}{\sigma^2} \left[ b^{**}+\frac{1}{2} (\beta-\mu^{**})^\top V^{**-1}(\beta-\mu^{**}) \right] \right\} d\beta d\sigma^2 \\
\end{align*}
where
\begin{align*}
\mu^{**} &= (V_{\beta}^{*-1} + X^{*\top} X^*)^{-1} (V_{\beta}^{*-1}\mu_{\beta}^* + X^{*\top} y^*), \\
V^{**} &= (V_{\beta}^{*-1} + X^{*\top} X^* )^{-1}, \\
b^{**} &= b^*  + \frac{1}{2}( \mu_{\beta}^{*\top} V_{\beta}^{*-1} \mu_{\beta}^*  + y_i^2- \mu^{**\top}V^{**-1}\mu^{**} ).
\end{align*}
Let $z\triangleq \frac{c}{\sigma^2}$, $c\triangleq b^{**}+\frac{1}{2} (\beta-\mu^{**})^\top V^{**-1}(\beta-\mu^{**})$, and then
\begin{align*}
f(y_i | \bm{y}) &= \frac{b^{*a^*}}{ (2\pi)^{\frac{p+1}{2}} \Gamma(a^*) |V^*|^{1/2}} \int c^{-(a^*+\frac{p+1}{2}+1)} z^{a^*+\frac{p+1}{2}+1} e^{-z} dcz^{-1} d\beta \\
&= \frac{b^{*a^*}}{ (2\pi)^{\frac{p+1}{2}} \Gamma(a^*) |V^*|^{1/2}} \int c^{-(a^*+\frac{p+1}{2})} d\beta \\
&= \frac{b^{*a^*}}{ (2\pi)^{\frac{p+1}{2}} \Gamma(a^*) |V^*|^{1/2}} \int \left[ b^{**}+\frac{1}{2} (\beta-\mu^{**})^\top V^{**-1}(\beta-\mu^{**}) \right] d\beta \\
&= \frac{b^{*a^*}}{ (2\pi)^{\frac{p+1}{2}} \Gamma(a^*) |V^*|^{1/2}} b^{**^{-\frac{2a^*+p+1}{2}}} \times \\
&\ \ \ \ \int  \frac{\Gamma(\frac{2a^*+p+1}{2})}{\Gamma(\frac{2a^*+1}{2}) \pi^{p/2}|(2a^*+1)[\frac{2b^{**}}{2a^*+1}V^{**}]|^{\frac{1}{2}}}  \left[ 1+\frac{(\beta-\mu^{**})^\top [\frac{2b^{**}}{2a^*+1}V^{**}] ^{-1}(\beta-\mu^{**})}{2a^*+1} \right]^{-\frac{2a^*+p+1}{2}}    d\beta .\\
\end{align*}
The integrable function above is the density of $\beta$, which is actually multivariate $t$-distribution MVS$t_{2a^*+1}(\mu^{**}, \frac{2b^{**}}{2a^*+1}V^{**})$. Since
$$|2b^{**}V^{**}|^{1/2} = 2^{p/2} b^{**\frac{1}{2}} \frac{|V^*|^{1/2}}{|1+X_iV^*X_i^\top|^{1/2}},$$
then we have
$$f(y_i | \bm{y}) = \frac{\Gamma(\frac{2a^*+1}{2}) }{ \sqrt{2a^*\pi} \Gamma(\frac{2a^*}{2}) [\frac{2b^*}{2a^*}(1+X_iV^*X_i^\top)]^{1/2}} \left[ 1+\frac{(y_i-X_i\mu^*)^2\{\frac{2b^*}{2a^*}(1+X_iV^*X_i^\top)\}}{2a^*} \right]^{-\frac{2a^*+1}{2}}.$$
It is like a univariate $t$-distribution shifted by $-X_i\mu^*$ and scaled by $\frac{\frac{2b^*}{2a^*}(1+X_iV^*X_i^\top)}{2a^*} $.

\cite{Vehtari2017} proposed the expected log pointwise predictive density (elpd) to measure the predictive accuracy for the new data $\tilde{y}_i$ ($i=1,\cdots,n$). The elpd is defined as 
$$\text{elpd}=\sum_{i=1}^n\int f(\tilde{y}_i)\log f(\tilde{y}_i |\bm{y}) d\tilde{y}_i.$$
Therefore, we use the log pointwise predictive density (lpd) to measure the predictive accuracy, i.e., lpd = $\sum_{i=1}^n \log f(y_i |\bm{y})$. The widely applicable information criterion (WAIC) is an alternative approach which is 
$$\text{lpd} - \sum_{i=1}^n \mathbbm{V}ar[\log f(y_i |\bm{y})].$$

\subsection*{Appendix 2: Posterior predictive for the dSUR and SSUR models}

For the dSUR and SSUR models, the response variables are independent in their reparametrised forms. It is feasible to use the out-of-sample predictive to measure the elpd. The Bayesian leave-one-out estimate is
$$\text{elpd}_{\text{loo}} = \sum_{i=1}^n \log f(y_i |\bm{y}_{-i}).$$
As note by the importance sampling, we get
$$f(y_i |\bm{y}_{-i}) \approx \frac{1}{\frac{1}{T}\sum_{t=1}^T \frac{1}{f(y_i |\bm{\theta}^t)} },$$
where all related parameters $\bm{\theta}^t$ are drawn from its full posterior. The WAIC is estimated by
$$\widehat{\text{elpd}}_{\text{waic}} = \widehat{\text{elpd}}_{\text{loo}} - \sum_{i=1}^n \mathbbm{V}ar_{t=1}^T[\log f(y_i |\bm{\theta}^t)].$$
The posterior predictive $f(y_i |\bm{y}_{-i})$ can be used to check outliers, which is also named the Conditional Predictive Ordinate (CPO, \cite{Gelfand1996}).


\end{document}